\documentclass[prd,nofootinbib,preprint,superscriptaddress]{revtex4}
\pdfoutput=1
\usepackage{hyperref}
\usepackage{amsmath,amssymb}
\usepackage{epsfig} 
\usepackage{graphicx}        
\usepackage{url}
\usepackage{color}
\usepackage{multirow}
\usepackage{placeins}
\usepackage[dvipsnames]{xcolor}
\usepackage{braket}
\usepackage{float}
\usepackage{slashed}
\usepackage{comment}
\hypersetup{
  colorlinks=true,
  linkcolor=red,
  filecolor=magenta,   
  urlcolor=violet,
  citecolor=blue,
}

\hypersetup{colorlinks,linkcolor={blue},citecolor={teal},urlcolor={violet}}
 
\allowdisplaybreaks

\bibpunct{[}{]}{,}{n}{}{,}

\def\beq{\begin{equation}}
\def\eeq{\end{equation}}
\def\bea{\begin{eqnarray}}
\def\eea{\end{eqnarray}}
\makeatletter

\newcommand{\documenttitle}{Cogenesis of baryon and dark matter with PBH and QCD axion}

\begin{document}

\title{\documenttitle}
\author{Debasish Borah}
\email{dborah@iitg.ac.in}
\affiliation{Department of Physics, Indian Institute of Technology Guwahati, Assam 781039, India}
\author{Nayan Das}
\email{nayan.das@iitg.ac.in}
\affiliation{Department of Physics, Indian Institute of Technology Guwahati, Assam 781039, India}
\author{Suruj Jyoti Das }
\email{surujjd@gmail.com}
\affiliation{Particle Theory  and Cosmology Group, Center for Theoretical Physics of the Universe,
Institute for Basic Science (IBS),
 Daejeon, 34126, Korea}
\author{Rome Samanta}
\email{romesamanta@gmail.com}
\affiliation{CEICO, Institute of Physics of the Czech Academy of Sciences, Na Slovance 1999/2, 182 21 Prague 8, Czech Republic}
\affiliation{Scuola Superiore Meridionale, Largo S. Marcellino 10, I-80138 Napoli, Italy}
\affiliation{  Istituto Nazionale di Fisica Nucleare (INFN), sez. di Napoli, Via Cinthia 9, I-80126 Napoli,
Italy}

\begin{abstract} 
With entropy injection, an early matter-dominated epoch (EMD) impels the axion decay constant $f_a$ towards larger values to produce correct axion dark matter (DM) abundance, thereby unfolding the low-mass axion ($m_a\lesssim 10^{-5}$ eV) parameter space to be searched for in axion experiments. We implement this proposition in a scenario where $f_a$ and the leptogenesis scale in a seesaw mechanism are equivalent. We show, that if instead, the EMD is provided by evaporating ultralight primordial black holes (PBH), the scenario becomes strikingly testable with gravitational waves (GW) background alongside the axion searches. In particular, while being consistent with correct axion DM abundance, the scale $f_a\gtrsim 10^{12}$ GeV, corresponding to the unflavored regime of leptogenesis with hierarchical right-handed neutrinos, can be probed with GW and axion experiments, which is otherwise not testable at neutrino or collider experiments. Additionally, axions produced from PBH evaporation can give rise to dark radiation within reach of future cosmic microwave background experiments.
\end{abstract}

\maketitle

\section{Introduction}
Strong upper limits on the electric dipole moment of neutron show that strong interactions are CP symmetric \cite{Abel:2020pzs}, although the standard model (SM) entirely is not. The CP violation in the strong interaction is characterized by the angle $\bar{\theta}$, which is bounded from above: $\bar{\theta}<10^{-10}$. The required vanishingly small value of the  $\bar{\theta}$ parameter is the origin of the strong-CP problem. The classic solution to this is to promote  $\bar{\theta}$ as a pseudoscalar axion field resulting from a global $U(1)$ symmetry breaking -- the  Peccei-Quinn (PQ) mechanism \cite{Peccei:1977hh, Peccei:1977ur, Wilczek:1977pj, Weinberg:1977ma}. When the axion field acquires its potential through QCD effects, remarkably, $\bar{\theta}$ relaxes to zero (ground state of the potential) and solves the strong-CP problem. Besides solving the strong-CP problem, axion could constitute the entire cold dark matter (DM) energy density if they are sufficiently light 
 $m_a\simeq 10^{-5}$ eV \cite{Preskill:1982cy, Abbott:1982af, Dine:1982ah}. The key UV parameter in any QCD axion model is the scale of the PQ symmetry breaking $f_a$, also known as the axion decay constant. In the models of light QCD axions,  $f_a$ is bounded from below by astrophysical observations as $f_a\gtrsim 10^{8}$ GeV \cite{Raffelt:2006cw, Caputo:2024oqc}. On the other hand, the vacuum misalignment mechanism offers a stringent window \footnote{Depending on the IR models, e.g., Dine-Fischler-Srednicki-Zhitnitsky (DFSZ) \cite{Dine:1981rt, Zhitnitsky:1980tq} and Kim-Shifman-Vainshtein-Zakharov (KSVZ) \cite{Kim:1979if, Shifman:1979if}, additional contributions from the topological defects can alter this window by less than an order of magnitude \cite{Kawasaki:2013ae}.} for the decay constant around $f_a\sim 10^{11}$ GeV to produce the correct dark matter relic \cite{Kawasaki:2013ae}. This simple PQ solution to the strong-CP problem thus renders dark matter physics and evades astrophysical constraints, generically for a high-scale phase transition corresponding to the PQ symmetry breaking. Since there exist other observed phenomena which the SM fails to explain, one might wonder whether any other cosmological puzzles attributed to high-scale physics in their simplest theoretical form share common ground with axion physics. 

This article discusses one such possibility, where the generation of the baryon asymmetry of the Universe (BAU) through thermal leptogenesis \cite{Fukugita:1986hr} shares a common origin with axion physics. While there are several ways in which one couples the axion physics to the baryogenesis \cite{Servant:2014bla, Ipek:2018lhm, Croon:2019ugf, Co:2019wyp}, our interest aligns with the SMASH-like \cite{Ballesteros:2016xej} scenarios (and similar extensions \cite{Clarke:2015bea, Sopov:2022bog}). In this scenario, the right-handed neutrinos (RHN) introduced in the SM to generate light neutrino mass via seesaw mechanism and to provide leptogenesis, get mass from the PQ phase transition, i.e., $f_a\equiv M\equiv \Lambda_{\rm lepto}$, where $M(\Lambda_{\rm lepto})$ is the RHN mass, equivalent to the scale of seesaw and thermal leptogenesis. Such a unification therefore addresses neutrino mass, strong-CP problem, DM, and BAU at one stroke.
\begin{figure*}
    \centering
    \includegraphics[width = 15 cm, height = 4 cm]{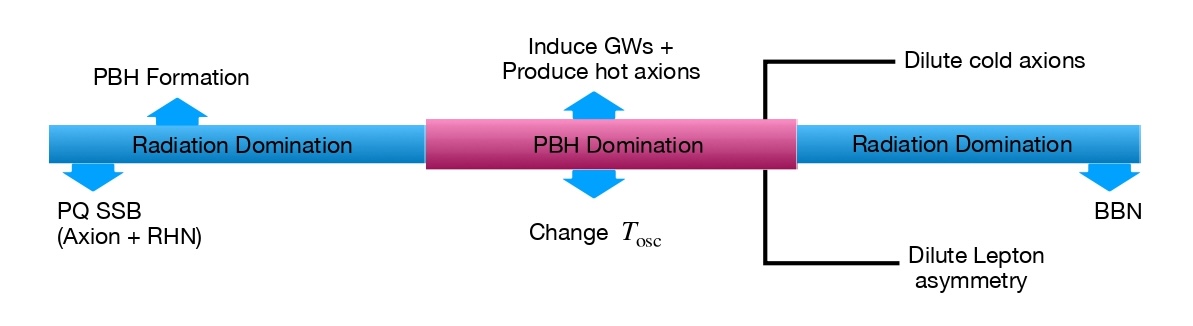}
    \caption{Schematic diagram showing a possible timeline of the scenario.}
    \label{fig:schematic}
\end{figure*}
Unlike Ref.\cite{Ballesteros:2016xej} for SMASH-like frameworks, the main motivation to unify the leptogenesis scale $\Lambda_{\rm lepto}$ and the PQ scale $f_a$ in this work is to explore the testability of $\Lambda_{\rm lepto}\equiv f_a\gtrsim 10^{12}$ GeV--also referred to as the vanilla regime of leptogenesis\footnote{Recently there have been some efforts to probe high-scale leptogenesis (including vanilla regime) with primordial gravitational waves e.g., topological defects \cite{Dror:2019syi, Blasi:2020wpy, Fornal:2020esl, Samanta:2020cdk, Barman:2022yos}.}. On the other hand, for $f_a\gtrsim 10^{12}$ GeV, axion dark matter gets overproduced via vacuum misalignment mechanism with standard cosmological expansion. In order not to lose the interesting connection among axion DM, strong-CP problem and vanilla leptogenesis, we explore the possibility that before the big bang nucleosynthesis (BBN), the Universe underwent an early matter-dominated phase releasing entropy while reproducing the standard radiation dominated Universe \cite{Visinelli:2009kt,Nelson:2018via}. 

In this work, we show that if the matter domination is provided by ultralight evaporating primordial black holes (PBH) \cite{Hawking:1974rv, Hawking:1975vcx,Carr:2020gox} with initial mass $m_\text{in}\lesssim 3.4\times 10^8\,\text{g}$, testability enhances significantly. This is because, in addition to predicting axion parameter space in the $g_{\alpha\gamma}-m_a$ plane, one can have stochastic gravitational waves (GWs) signatures via PBH density fluctuations \cite{Papanikolaou:2020qtd, Domenech:2020ssp, Domenech:2021wkk, Domenech:2021ztg, Papanikolaou:2022chm}. Remarkably, the mass range of PBH required to generate correct DM and at the same time BAU, produces GWs within the frequency bands of current and planned GW detectors.  A schematic diagram of this scenario showing a possible timeline is presented in Fig. \ref{fig:schematic}. Two additional (not mentioned so far) features in this figure namely,  the axion oscillation temperature $T_{\rm osc}$ and the production of hot axions in the form of dark radiation within reach of future cosmic microwave background (CMB) experiments are discussed later in this article.

Let us also add the following remarks. 

$\bullet$  Production of axion DM with PBH-induced matter domination has been studied in Ref.\cite{Bernal:2021yyb, Mazde:2022sdx}. The crucial difference here nonetheless is that, because we implemented the PBH-induced matter domination in SMASH-like scenarios, the requirement of generating the observed baryon asymmetry restricts the model parameter space severely to predict constrained GW-signal in amplitude as well as in frequency (see, e.g., Eq.\eqref{eq:betaM1}, \eqref{eqn:omgpeak_hie}, and Fig.\ref{fig:GW_SPEC}). As such, a hierarchical (resonant) leptogenesis scenario would predict weaker (stronger) GWs with detectable peak frequencies varying within the mHz-Hz range. 

$\bullet$ Given the early Universe before BBN epoch is unknown, the motivation to introduce PBH-induced pre-BBN matter domination in this work is purely phenomenological; as recently explored in a plethora of works, see, e.g., Refs.\cite{Gondolo:2020uqv, Bernal:2020bjf, Green:1999yh, Khlopov:2004tn, Dai:2009hx, Allahverdi:2017sks, Lennon:2017tqq, Hooper:2019gtx, Sandick:2021gew, Fujita:2014hha, Datta:2020bht, JyotiDas:2021shi,Barman:2021ost, Barman:2022gjo, Barman:2022pdo, Cheek:2021odj, Chaudhuri:2023aiv, Borah:2024lml,Bernal:2021yyb, Mazde:2022sdx} related to DM models, and Refs. \cite{Hawking:1974rv, Carr:1976zz, Baumann:2007yr, Hook:2014mla, Fujita:2014hha, Hamada:2016jnq, Morrison:2018xla, Hooper:2020otu, Perez-Gonzalez:2020vnz, Datta:2020bht, JyotiDas:2021shi, Smyth:2021lkn, Barman:2021ost, Bernal:2022pue, Ambrosone:2021lsx} for baryogenesis models. While PBH might arise owing to different (independent of the framework discussed here) new physics \cite{Carr:2020gox,Franciolini:2021nvv,Yoo:2022mzl,Bhattacharya:2023ztw}, a self-consistent framework would be to explore the formation of ultralight PBH in this model itself, e.g., via domain wall collapse \cite{Ferrer:2018uiu}. We, however, do not explore such a possibility here. 

The rest of the paper is organized as follows. In sec.\ref{sec2}, we briefly discuss the particle physics framework we have adopted in this work. The interplay of axion DM and leptogenesis in the presence of PBH is presented in sec.\ref{sec3}. In sec.\ref{sec4}, we discuss the detection prospect of the framework in GW and axion experiments. We conclude in sec.\ref{sec6}.
\section{The Framework}
\label{sec2}
We consider a type-I seesaw framework \cite{Minkowski:1977sc, GellMann:1980vs, Mohapatra:1979ia, Yanagida:1980xy, Schechter:1980gr} extended by Peccei-Quinn symmetry \cite{Peccei:1977hh, Peccei:1977ur}. A recent review of QCD axion models can be found in \cite{DiLuzio:2020wdo}. In addition to the heavy RHNs (here we consider three RHNs) required for the seesaw, the other field content and PQ charges are kept similar to a KSVZ type model \cite{Kim:1979if, Shifman:1979if} in which $v_{\rm PQ}=f_a$, where $v_{\rm PQ}$ is the vacuum expectation value (VEV) of the PQ scalar field $\sigma$ and $f_a$ is the axion decay constant\footnote{It should be noted that our particle content is same as the SMASH framework \cite{Ballesteros:2016xej}. While SMASH considers Higgs portal inflation via non-minimal coupling to gravity, we remain agnostic about the origin of inflation.}. Thus, the BSM field content is constituted by a complex PQ scalar field $\sigma \equiv \dfrac{v_{\rm PQ} + \rho}{\sqrt{2}}\, e^{i {a}/{f_{a}}}$ ($\sigma \sim (1,1,0)$), a heavy quark $Q \sim (3,1,0)$ and three right-handed neutrinos $N_{R} \sim (1,1,0)$ where the numbers in brackets denote their charges under the SM gauge symmetry $SU(3)_c \times SU(2)_L \times U(1)_Y$. The transformation of these fields and leptons under PQ symmetry are 
\begin{align}
     \sigma \, \rightarrow \, e^{i\alpha}\,  \sigma; \,\,\,\,
    N_{R} \, \rightarrow \, e^{-i{\alpha}/{2}} \, N_{R}; \,\,\,\,  l_{R}, L \, \rightarrow \, e^{-i{\alpha}/{2}}\,  l_{R}, L; \,\,\,\, 
       Q_{L} \, \rightarrow \, e^{i{\alpha}/{2}} \, Q_{L}; \,\,\,\, Q_{R} \, \rightarrow \, e^{-i{\alpha}/{2}} \, Q_{R}. \nonumber 
\end{align}
The SM quarks have vanishing PQ charges. The Yukawa coupling part of the PQ invariant Lagrangian can be written as
\begin{equation}\label{yukawalag}
    \mathcal{L_Y} \, = - \left[ \, y \overline{Q}_L \sigma Q_R \, + \, G_{ij} \overline{L_i}H l_{jR} \, + \, F_{ij} \overline{L_i} \Tilde{H} N_{jR} \, + \, \dfrac{1}{2}\, y_{ij} \, \overline{N}_{iR}^c\sigma N_{jR} \, \right] + \, {\rm h.c.}
\end{equation}
with $H$ being the SM Higgs. The scalar potential of the model is given by 
\begin{equation}
    V(H, \sigma) \, = \, \lambda_H \left(H^\dagger H \, - \, \dfrac{v^2}{2}\right)^2 \, + \, \lambda_\sigma \left(|{\sigma}|^2\, - \, \dfrac{v_{\rm PQ}^2}{2}\right)^2 \, +\, \lambda_{H \sigma} \left(H^\dagger H \, - \, \dfrac{v^2}{2}\right) \, \left(|{\sigma}|^2\, - \, \dfrac{v_{\rm PQ}^2}{2}\right)\, 
\end{equation}
with $v $ being the VEV of the SM Higgs $H$.
After PQ symmetry breaking the first and the last terms in $\mathcal{L_Y}$ give rise to
\begin{eqnarray}
     y \bar{Q}_L \sigma Q_R \, \rightarrow \,  \frac{y}{\sqrt{2}} \rho \bar{Q}_L Q_R  e^{i {a}/{f_{a}}} \, + \, \frac{y}{\sqrt{2}} v_{PQ} \bar{Q}_L Q_R  e^{i {a}/{f_{a}}}  \\ \nonumber
     \dfrac{1}{2}\, y_{ij} \, \Bar{N}_{iR}^c\sigma N_{jR} \, \rightarrow \,  \dfrac{1}{2}\, \frac{y_{ij}}{\sqrt{2}} \rho \, \Bar{N}_{iR}^c N_{jR} e^{i {a}/{f_{a}}} \, + \, \dfrac{1}{2}\, \frac{y_{ij}}{\sqrt{2}} v_{PQ} \, \Bar{N}_{iR}^c N_{jR} e^{i {a}/{f_{a}}}.
\end{eqnarray}
The phase part can be absorbed by the transformation $ Q_{R} \, \rightarrow \, e^{-i\frac{a}{2 f_a}} Q_R$ and $N_{R} \, \rightarrow \, e^{-i\frac{a}{2 f_a}} \, N_{R}$. As the chiral transformation on $Q$ is anomalous under QCD, this gives the term $\frac{g^2_s}{32 \pi^2} \frac{a}{f_a} G_{\mu \nu} \Tilde{G}^{\mu \nu}$ with $G_{\mu \nu} \, (\Tilde{G}^{\mu \nu})$ being (dual) field strength tensor of QCD. Simultaneously, from the kinetic term $\overline{Q}i\gamma^{\mu}\partial_{\mu} Q$, after transformation one gets the term $-\frac{\partial_{\mu} a}{2 f_{a}}\overline{Q}\gamma^{\mu} \gamma_{5} Q$. Similarly, from the kinetic term of RHN $\dfrac{i}{2} \overline{N} \gamma^\mu \partial_\mu N$, one gets the term $-\dfrac{1}{4} \, \dfrac{ \partial_\mu a }{ \, f_a} \, \overline{N} \gamma^\mu \gamma_5 N$. Now using the Dirac equation and the fact that the total derivative is zero at the boundary, we get 
\begin{eqnarray}
    -\frac{\partial_{\mu} a}{2 f_{a}}\overline{Q}\gamma^{\mu} \gamma_{5} Q \, = \,  i \frac{M_Q}{f_{a}} a \overline{Q} \gamma_{5} Q \nonumber \\
    -\dfrac{1}{4} \, \dfrac{ \partial_\mu a }{ \, f_a} \, \overline{N} \gamma^\mu \gamma_5 N \, = \,  \dfrac{i}{2} \, \dfrac{M_i}{f_a} \, a \, \overline{N} \gamma_5 N
\end{eqnarray}
which define axion couplings to RHN and heavy quark $Q$. We have also dropped the chirality index from RHN notation and will denote them only by $N_i$ with mass $M_i$ hereafter.

\section{Axion dark matter,  PBH, and Leptogenesis}
\label{sec3}
 We consider PBH to be formed in the early radiation-dominated Universe at a temperature $T_{\rm in}$, with an initial mass related to the particle horizon size as \cite{Fujita:2014hha,Masina:2020xhk}
\begin{align}
    m_{\rm in}=\frac{4}{3}\,\pi\,\gamma\,\Bigl(\frac{1}{\mathcal{H}\left(T_\text{in}\right)}\Bigr)^3\,\rho_\text{R}\left(T_\text{in}\right)\,
\label{eq:pbh-mass}
\end{align}
with $\rho_\text{R}$, $\mathcal{H}$ indicating the radiation energy density and the Hubble parameter respectively. $\gamma\simeq 0.2$ is a numerical factor that contains the uncertainty of the PBH formation. The initial abundance of PBH during their formation is denoted by
\begin{align}
     \beta = \frac{\rho_{\rm PBH}\left(T_\text{in}\right)}{\rho_{\rm R}\left(T_\text{in}\right)}\,,
\end{align}
where $\rho_{\rm PBH}$ represents the PBH energy density. After their production, PBH evolve as matter and dominate the energy density of the early Universe, if their initial fractional energy density is greater than a critical value given by
\begin{align}
    \beta \geq \beta_{\text{crit}} \simeq 2.5\times 10^{-14} \gamma^{-\frac{1}{2}} \left(\frac{m_{\rm in}}{10^8 \text{g}}\right)^{-1}\,. \label{eq:betacr}
\end{align}
Finally, PBH lose their mass through Hawking radiation and completely evaporate at a temperature 
\begin{align}
      T_{\rm ev}\simeq\left(\frac{9g_{*,B}(T_{\rm BH})}{10240}\right)^{\frac{1}{4}}\left(\frac{M_{\rm P}^{5}}{m_{\rm in}^{3}}\right)^{\frac{1}{2}}\,,\label{eq:Tev}
\end{align}
where $M_{P}$ is the reduced Planck mass and $g_{*,B}(T_{\rm BH})\simeq 100$ being the relativistic degrees of freedom below $T_{\rm BH}$. For QCD axions, the zero temperature mass ($T\leq T_{\rm QCD} \simeq$ 150 MeV) is related to
PQ symmetry breaking scale, $f_{a}$ as
\begin{align}
    m_{a} \simeq 5.7 \left(\frac{10^{12}~\rm GeV}{f_{a}}\right) \rm \mu eV. 
\end{align}
Above $T>T_{\rm QCD}$, the temperature-dependent axion mass is given as 
\begin{align}
    m_{a} (T) = m_{a} \left(\frac{T_{\rm QCD}}{T}\right)^{4}.
\end{align}
As indicated by lattice simulation, the power of 4 is not precise \cite{DiLuzio:2020wdo}, but it does not affect our overall results significantly. If the PQ symmetry breaks in post inflationary era, the initial misalignment angle, $\theta_{i}= a_{i}/f_{a}$ takes the average value $\theta_{i} = \frac{\pi}{\sqrt{3}} \sim 1.81$. At high temperatures, the axion evolution is friction-dominated and frozen at $\theta_i$. It then starts to oscillate at a temperature $T_{\rm osc}$, when the Hubble rate becomes comparable to axion mass: $\mathcal{H} (T_{\rm osc}) \sim 3 m_{a}(T_{\rm osc})$. From the onset of oscillations temperature, axion starts behaving as matter, and from conservation of comoving number density, its number density at a later epoch can be written as 
\begin{align}
    n_{a}(T) = n_{a}(T_{\rm osc}) \frac{s(T)}{s(T_{\rm osc})},
\end{align}
where $s(T)$ is the comoving entropy density at a temperature $T$. The axion behaves as cold matter and can make all the observed dark matter abundance with total abundance 
\begin{align}
    \Omega_{a}h^2 = \frac{\rho_{a}(T_{0})}{\rho_{c}}h^2 = \frac{m_{a}(T_{0})}{\rho_{c}} \left(\frac{\rho_{a}(T_{\rm osc})}{m_{a}(T_{\rm osc})} \frac{s(T)}{s(T_{\rm osc})} \right) h^2.
\end{align}
Here $\rho_{\rm c}$ represents the critical energy density of the Universe today. If axion constitutes all the dark matter, it should have a mass around $\sim 14 $ $\mu$eV, when produced via the misalignment mechanism \cite{DiLuzio:2020wdo}. The presence of PBH modifies the axion dark matter abundance in two different ways:  $(1)$ PBH change the oscillation temperature $T_{\rm osc}$, because $\rho_{\rm PBH}$ contributes to Hubble expansion rate $\mathcal{H}$; $(2)$ PBH evaporation dilutes the existing abundance of axions via entropy injection. The latter effect is the dominant one. In addition to affecting the axion dark matter abundance, PBH evaporation can also produce axions. However, these axions behave as dark radiation contributing to $N_{\rm eff}$. We numerically study the evolution of the PBH-axion system, closely following Ref. \cite{Mazde:2022sdx} (see Appendix\ref{appen1}).

Lepton asymmetry can be generated thermally (with the thermal production of RHNs mediated by Yukawa couplings) at a high scale via the CP-violating decays of RHNs, which can be converted to the baryon asymmetry through the electroweak sphaleron processes \cite{Fukugita:1986hr}. At a temperature $T\sim M_i$, lepton asymmetry produced by $i$th RHN freezes out; therefore, $M_i$ is typically the scale of thermal leptogenesis; $M_i\equiv \Lambda_{\rm lepto}$. In the high-scale leptogenesis scenario, there are three distinct regimes of leptogenesis. The RHNs produce lepton doublets as a coherent superposition of flavors:  $\ket{L_i} = A_{i \alpha }\ket{L_{\alpha i}}$, where $\alpha = e,\mu,\tau$ and $ A_{i\alpha }$ is the corresponding amplitude determined by $F_{i\alpha}$.  For temperatures $T\gtrsim 10^{12}$ GeV, all the charged lepton Yukawa couplings are out of equilibrium (the interaction strength is given by $\Gamma_\alpha\sim 5\times 10^{-3} T h_\alpha $, with $h_\alpha$ being the charged lepton Yukawa coupling) and do not participate in the leptogenesis process.  This is the unflavor or the vanilla regime of leptogenesis. In this regime, leptogenesis is not sensitive to the neutrino mixing matrix, e.g., mixing angles and low-energy CP phases \cite{Pascoli:2006ci}.  When the temperature drops down to $T\lesssim10^{12}$ GeV, the $\tau$ flavored charged lepton interaction comes into equilibrium and breaks the coherent $\ket{L_i}$ state into $\tau$ and a composition of $e+\mu$ flavor (two flavor regime). Likewise, the later composition is broken once $T\lesssim 10^9$ GeV, when the $\mu$ flavored charged lepton interaction comes into equilibrium (three flavor regime). In both these flavored regimes, leptogenesis may be sensitive to the low-energy neutrino parameters because the neutrino mixing matrix cannot be eliminated from the CP asymmetry parameter. Therefore, for $T\sim M_i\gtrsim 10^{12}$ GeV, practically there are no ways to test high-scale thermal leptogenesis, barring some recent effort with gravitational waves from topological defects \cite{Dror:2019syi, Blasi:2020wpy, Fornal:2020esl, Samanta:2020cdk, Barman:2022yos}. As mentioned in the introduction, here we show, in the unified setup of axion and seesaw, the vanilla regime is testable with axion physics as well as gravitational waves from PBH density fluctuations. 

A crucial aspect of this work is to obtain the correct DM density for large $f_a\gtrsim 10^{12}$ GeV, requiring PBH evaporation after axion oscillation so that the overdensity of DM can be diluted. As we shall see, the required range of PBH mass for this does not affect the lepton asymmetry production (happens around $T\sim f_a$), however, dilutes it similarly to the DM density, and therefore correlates both of them. 

 For hierarchical RHNs,  the CP asymmetry parameter can be estimated as \cite{Buchmuller:2004nz, Davidson:2008bu}  
\begin{equation}\label{eq:epsilon}
    \epsilon_{1} = \frac{3 M_{1} \sqrt{(\Delta m_{\rm atm})^2}}{4 \pi v^2},
\end{equation}
with $v/\sqrt{2}=174$ GeV is the VEV of the SM Higgs and $(\Delta m_{\rm atm})^2 \simeq 2.4\times 10^{-3} \rm eV^{2}$ is the active neutrino atmospheric mass-squared difference. The resulting baryon asymmetry from the decay of $N_{1}$ can be approximately written in terms of baryon-to-photon ratio as \cite{Borah:2022iym}
\begin{equation}\label{eq:eta_B}
    \eta_{B} \approx 10^{-2} \kappa_{1} \frac{\epsilon_{1}}{\xi}\,.
\end{equation}
Here $\xi$ denotes the entropy dilution due to PBH evaporation, estimated as \cite{Borah:2022iym}
\begin{equation}\label{eq:xi}
    \xi = 233 \, \beta \left(\frac{m_{\rm in}}{M_{\rm Pl}}\right) \left(\frac{\gamma}{g_{*,B}(T_{\rm BH})\mathcal{G}}\right)^{1/2},
\end{equation}
and $\kappa_1\simeq 10^{-2}$ is the efficiency of lepton asymmetry production. $\mathcal{G}\sim 3.8$ is the greybody factor and $M_{\rm Pl}= \sqrt{8\pi} M_{\rm P}$ is the Planck mass. The extra factor $10^{-2}$ on the right-hand side of Eq.\eqref{eq:eta_B} accounts for the combined effects of sphaleron conversion of the leptons to baryons plus the photon dilution \cite{Buchmuller:2004nz}. Because of our proposed unification $f_{a} \simeq M_{1}$,\footnote{Note that here we use the formula for the CP asymmetry parameter valid only for a hierarchical mass spectrum of RHNs, and we consider a $N_1$-dominated leptogenesis scenario. In this case, although the heavier RHNs do not have any significant role to play, their masses are constrained as $M_{i=2,3}<f_a\sqrt{4\pi}$.}, Eq.\eqref{eq:epsilon}, \eqref{eq:eta_B}, and \eqref{eq:xi} impose the following condition on $\beta$:
\begin{align}
    \beta\simeq5.7\times 10^{-12}\left(\frac{M_{\rm Pl}}{m_{\rm in}}\right) \frac{f_a}{{\rm GeV}}\simeq 32.3\left(\frac{M_{\rm Pl}}{m_{\rm in}}\right) \left(\frac{1\rm \mu eV}{m_{a}}\right)\,.\label{eq:betaM1}
\end{align}
Eq.\eqref{eq:betaM1} is the most important analytical relation we derive. Because of this stringent relation among $\beta$, ${m_{\rm in}}$, and $f_a(m_a)$, concurrent generation of correct DM and BAU occurs for an extremely constrained parameter space as analyzed below in detail. 

\begin{figure}
    \centering
    \includegraphics[scale=0.5]{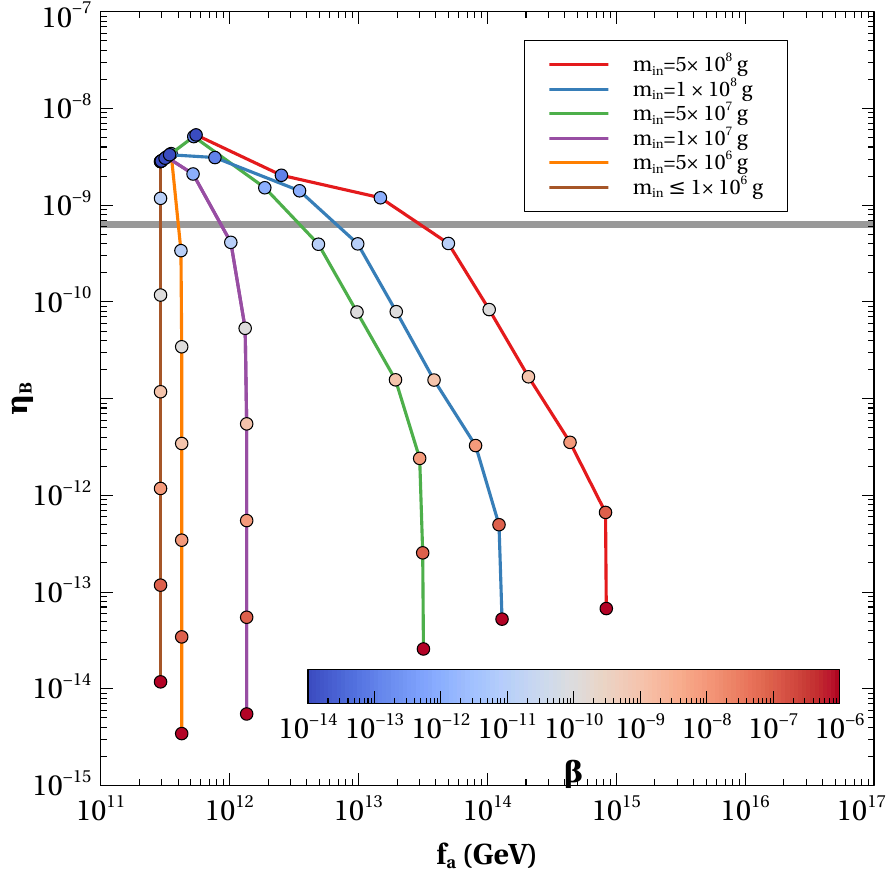}
    \caption{Baryon to photon ratio, $\eta_{B}$ versus $f_{a}$ for different PBH mass and initial fractional abundance, $\beta$. Here, the contours satisfy the correct DM relic. Initial misalignment angle is taken to be $\theta_{i}=\frac{\pi}{\sqrt{3}}\sim 1.81$. The horizontal gray line denotes the observed value of $\eta_{B}$. Here, we consider $M_1 = f_a$ which gives us Eq.\eqref{eq:betaM1}.}
    \label{fig:fa_vs_asymmetry}
\end{figure}
In Fig.\ref{fig:fa_vs_asymmetry}, we show the baryon asymmetry as a function of the PQ scale $f_a$, for different values of initial PBH mass. The contours shown are consistent with the observed DM relic from axion oscillation. If PBH evaporates before the axion oscillation temperature $T_{\rm osc}$, they do not affect the axion dynamics. For $m_{\rm in} \lesssim 10^{6}$ g, the PBH evaporation temperature remains higher than $T_{\rm osc}$ leading to the standard axion results with $f_a \simeq 3\times 10^{11}$ GeV, generating the correct DM relic.  This is indicated by the left-most brown-colored contour in Fig.\ref{fig:fa_vs_asymmetry}. Along this contour, $\beta$ is varied, as per the color bar, which changes the baryon asymmetry because of entropy dilution, following Eq.\eqref{eq:xi}. Increasing the PBH mass above $10^{6}$ g decreases the oscillation temperature and reduces axion abundance because of entropy dilution, resulting in a higher $f_a$ to obtain the observed DM abundance. With a decrease in $\beta$, this cumulative effect of PBH starts to diminish and hence the contours move towards the standard case of $f_{a}\simeq 3\times 10^{11}$ GeV. 

On the other hand, decreasing $f_a$ and $\beta$ has opposite effects on the baryon asymmetry $\eta_B$, with the former decreasing the asymmetry by decreasing $M_1$ (cf. Eq.\eqref{eq:epsilon}), while the latter increasing the asymmetry because of lesser entropy dilution. The latter effect turns out to be more dominant, leading to an increase in the value of $\eta_B$.

The lepton asymmetry can be significantly enhanced if two of the RHNs are quasi-degenerate, i.e., $\Delta M=M_2-M_1\ll \overline{M}=(M_1+M_2)/2$ leading to the resonant leptogenesis scenario \cite{Pilaftsis:2003gt, Dev:2017wwc}. For $\Delta M \simeq \Gamma$, the CP asymmetry parameter can reach up to $\mathcal{O}(1)$, where $\Gamma$ indicates the decay width of the RHN. For the resonant scenario, the expression for $\beta$ can be derived as 
\begin{eqnarray}
    \beta = 2.9\times10^{4} \epsilon_{1} \left(\frac{M_{\rm Pl}}{m_{\rm in}}\right). \label{eq:beta_reso}
\end{eqnarray}
We quantify the results for both, the hierarchical and the resonant leptogenesis scenarios on the  $m_{\rm in}-\beta$ plane 
\begin{figure*}
    \includegraphics[scale=0.42]{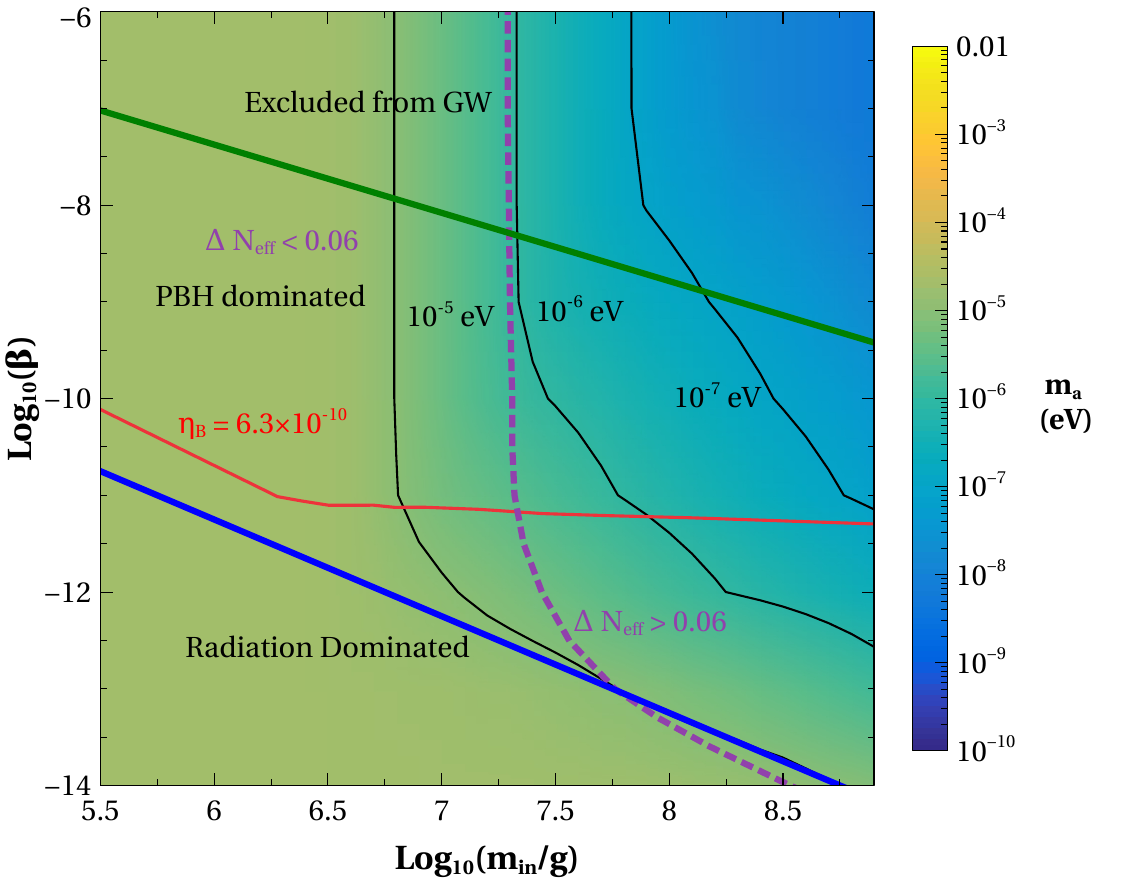}
    \includegraphics[scale=0.42]{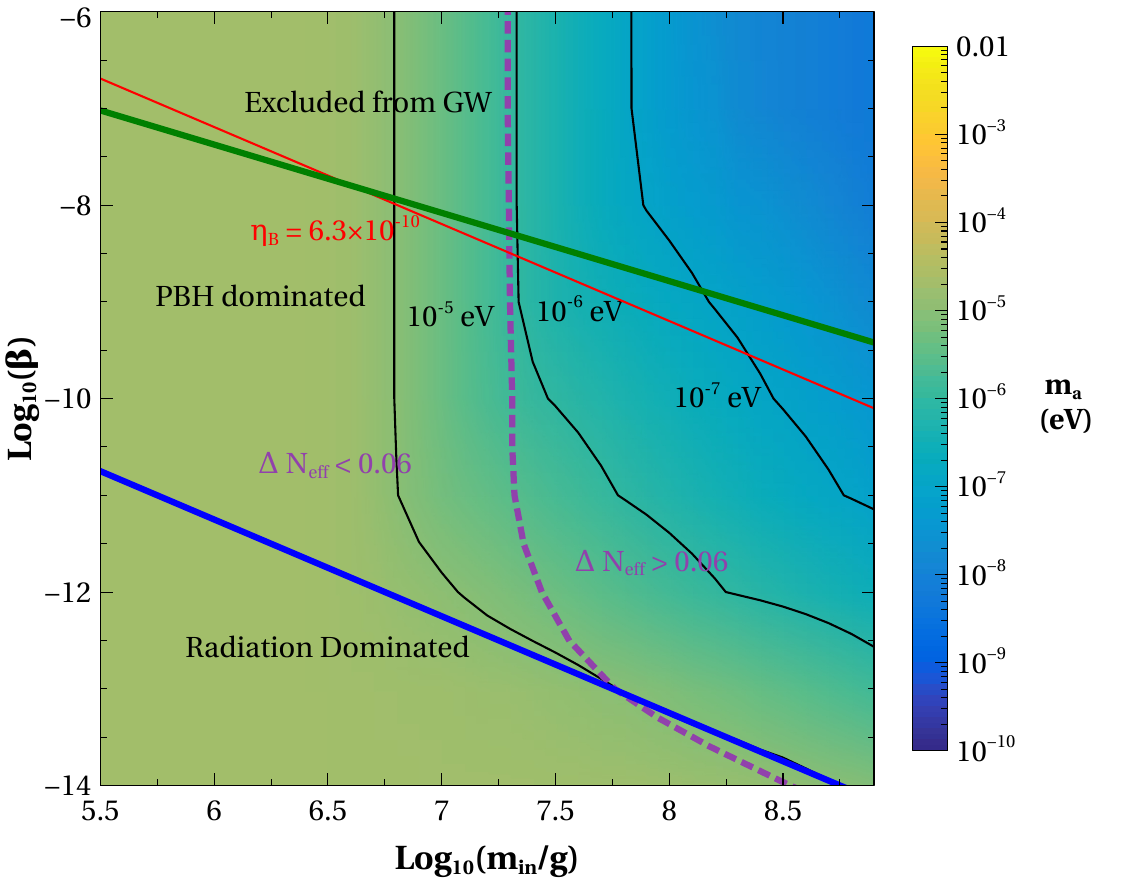}
    \caption{Constraint on $m_{\rm in}$ versus $\beta$ plane considering a hierarchical $N_1$-dominated leptogenesis scenario (left plot) and a resonant leptogenesis scenario (right plot). The color bar represents the axion mass that satisfies the observed DM abundance. The red color lines in both plots denote the parameter space that is consistent with both, DM relic and baryon asymmetry. The region excluded from GW is separated by the solid green line. The magenta dashed line separates the regions that can be probed by future CMB-S4 experiments due to the contribution of hot axion on $N_{\rm eff}$.}
    \label{fig:mbh_vs_beta}
\end{figure*}
in Fig.\ref{fig:mbh_vs_beta}. The variation of axion mass is represented by the color gradient (see also the contours).  For PBH mass $m_{\rm in}\lesssim 10^{6.5}$ g, the evaporation of PBH occurs before the onset of axion oscillation, thereby not changing the axion abundance. Similarly, in the radiation domination region (below the blue line), PBH do not affect the axion abundance.  For  $m_{\rm in}\gtrsim 10^{6.5}$ g, PBH delay the oscillation temperature plus dilute the axion density. Therefore, an ample amount of axions is required before PBH evaporation so that the correct DM relic is satisfied. This is achieved by reducing the axion mass (or increasing the PQ symmetry breaking scale $f_{a}$). The larger the $m_{\rm in}$, the stronger the entropy production and therefore smaller the axion mass (larger $f_a$).

The observed baryon asymmetry constraints for the hierarchical (left) and resonant scenario (right) are shown with the red curves. For the PBH mass range, $10^{5.5} \rm g \lesssim m_{\rm in} \lesssim 10^{6.5} \rm g$, the CP asymmetry parameter remains constant as $M_1 \sim f_a\sim 10^{11}$ as the axion DM production is not affected by PBH. Therefore, for a fixed $f_a$, the correlation between $\beta$ and $m_{\rm in}$ determined according to Eq.\eqref{eq:betaM1} ($\beta$ decreases with the increase of $m_{\rm in}$). For $m_{\rm in} \gtrsim 10^{6.5}$ g, an increase in PBH mass increases the axion scale $f_{a}$ owing to the interplay of entropy dilution and producing correct DM abundance. Because the  CP asymmetry increases monotonically with $f_a$, in this case, a sharp decrease in $\beta$ with the increase of $m_{\rm in}$ gets partially compensated, resulting in a plateau in the red curve (left). The region below (above) the red curve corresponds to $\eta_{B} >(<)\, 6.3\times10^{-10}$. As we shall discuss in sec.\ref{sec4}, the allowed values of $\beta$ in this hierarchical leptogenesis scenario produce GWs with weaker amplitude (the maximum value of $\beta$ allowed by $N_{\rm eff}$ constraint on GWs is shown with the green line) compared to its resonant version. 

In the right panel of Fig.\ref{fig:mbh_vs_beta}, we show the results for the resonant leptogenesis scenario. Instead of Eq.\eqref{eq:epsilon}, here we have considered a maximal CP asymmetry $\epsilon_{1} = 0.1$, achievable with resonant enhancement. Due to this enhancement in the value CP asymmetry parameter value, strong entropy dilution and therefore for a given PBH mass, a larger value of $\beta$  is required to produce the observed BAU. This is evident from the upward shift of the solid red line compared to the one in the left panel. In addition, due to a larger value of $\beta$, the resonant leptogenesis scenario produces much stronger GWs compared to the hierarchical case. 

For completeness, in both the plots, we also indicate the parameter space within reach future CMB experiments like CMB-S4 \cite{Abazajian:2019eic} which can measure additional relativistic degrees of freedom $\Delta N_{\rm eff}$. Such non-zero $\Delta N_{\rm eff}$ or dark radiation arises in our setup due to the production of hot axions from PBH evaporation. Note remarkably that in both the plots, the $T\sim M_i\sim f_a\gtrsim 10^{12}$ GeV is potentially testable, with either axion detectors or in combination with GW experiments.

\section{Detection prospects}
\label{sec4}
\subsection{Signatures in the form of GW from PBH density fluctuation }
\begin{figure*}
    \centering
    \includegraphics[width=8 cm, height=6 cm]{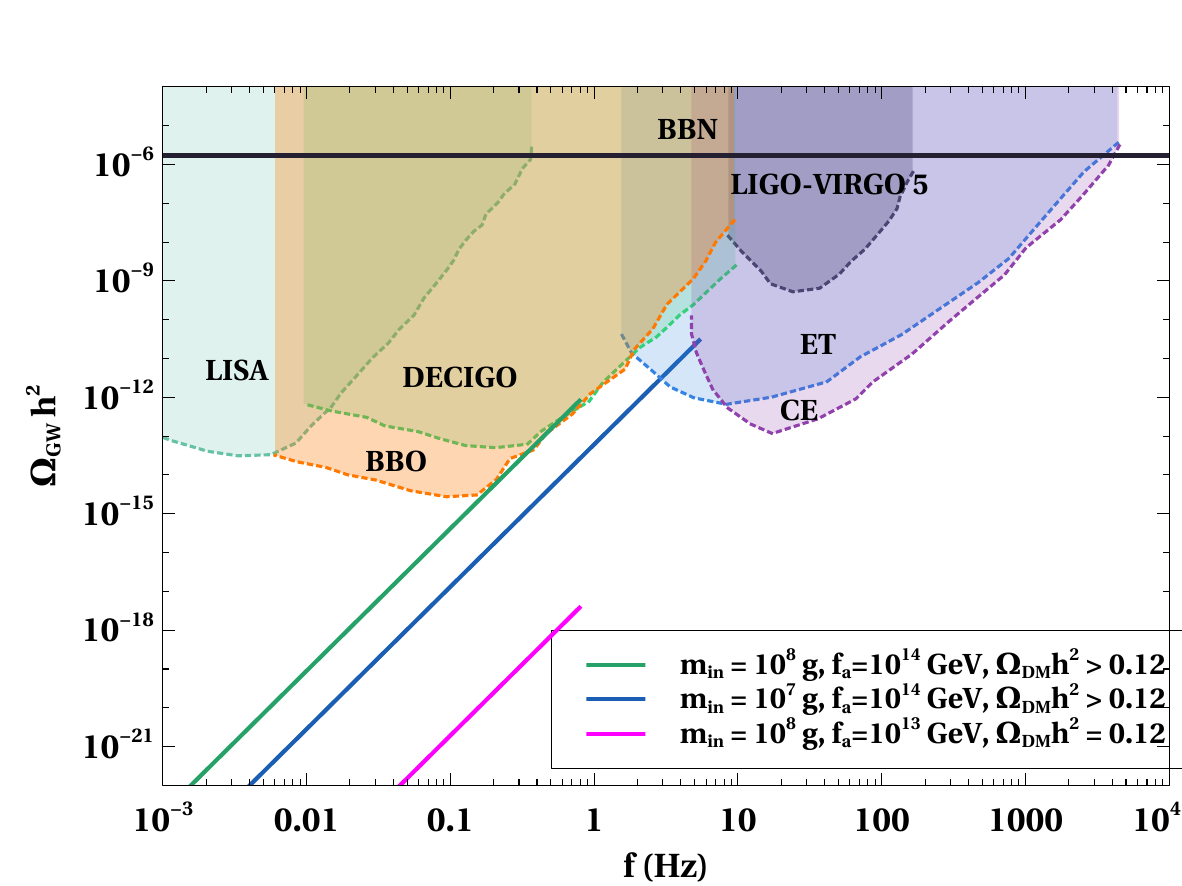}
     \includegraphics[width=8 cm, height=6 cm]{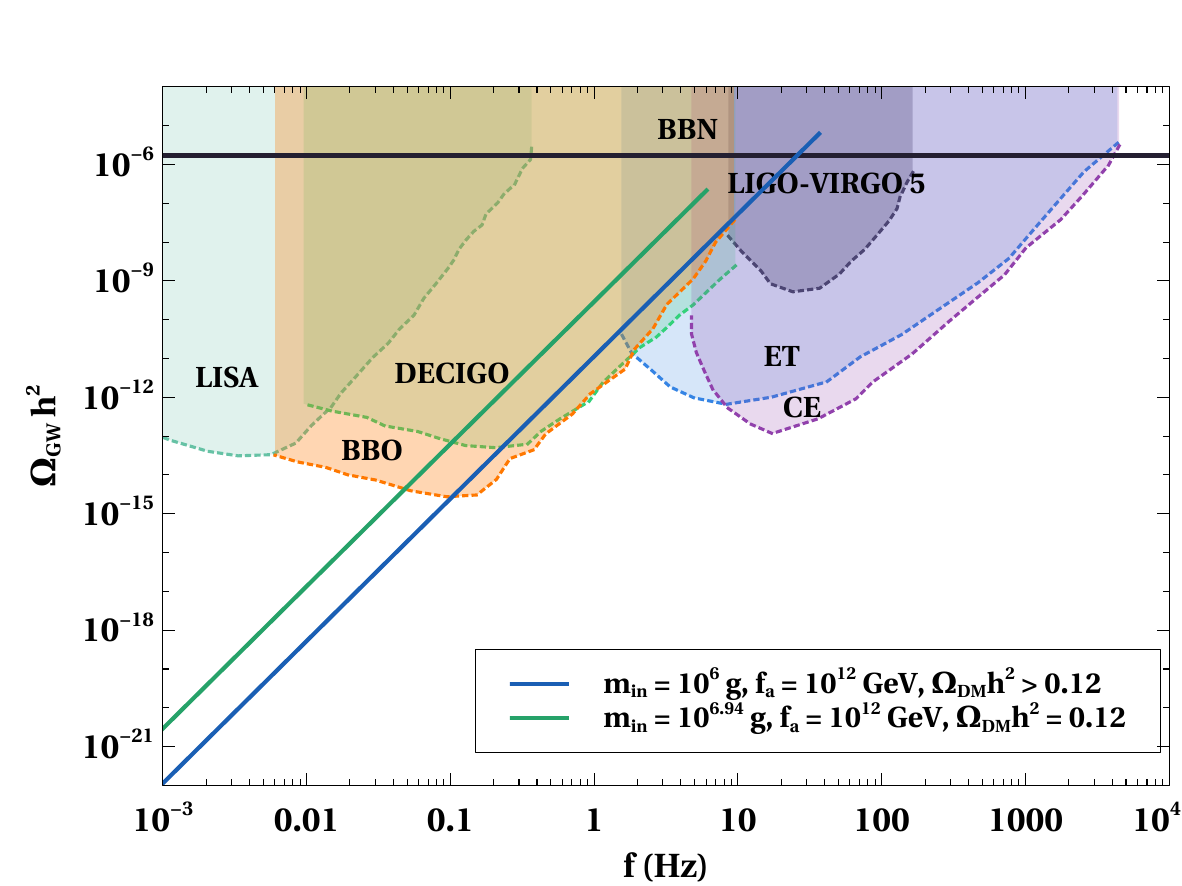}
    \caption{Left panel: GW spectrum for hierarchical scenario. Right panel: GW spectrum for the resonant scenario.}
    \label{fig:GW_SPEC}
\end{figure*}
\begin{figure*}
    \centering
    \includegraphics[width=8 cm, height=6 cm]{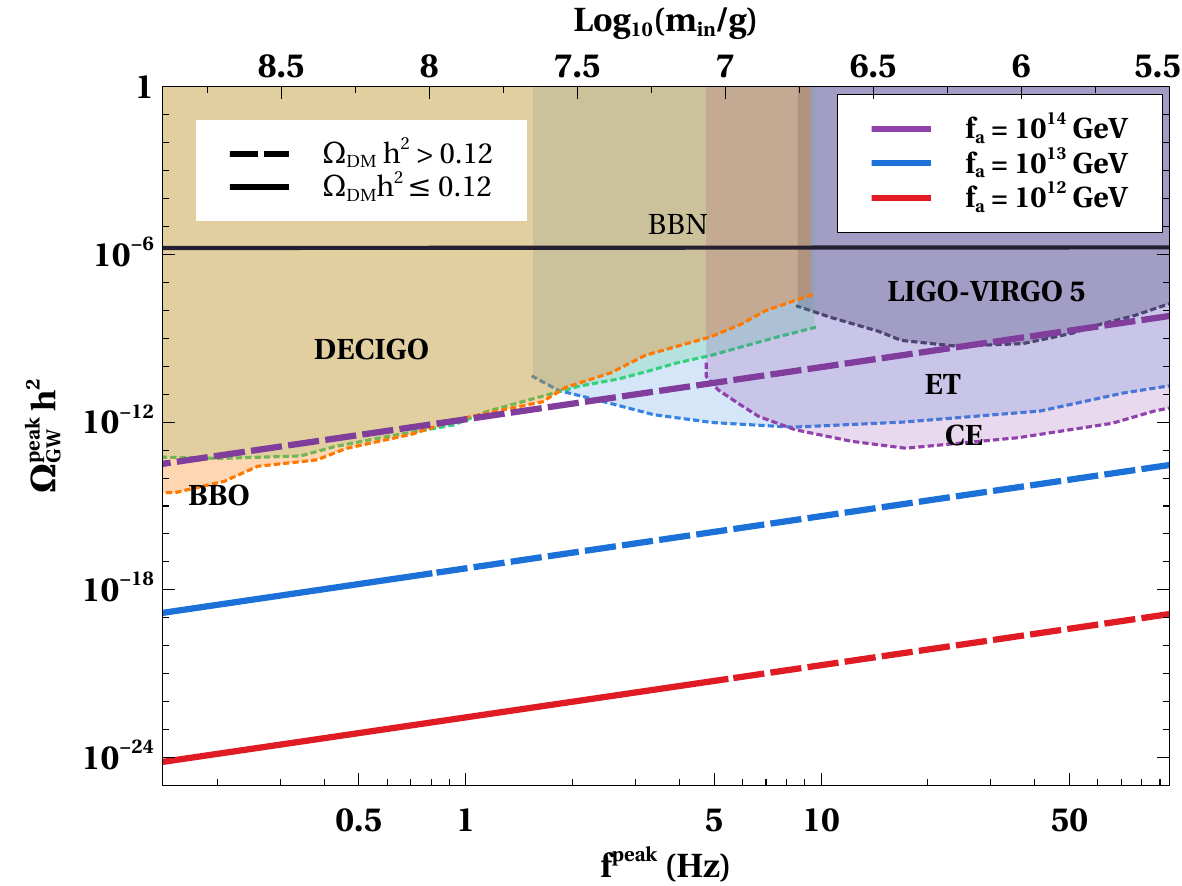}
     \includegraphics[width=8 cm, height=6 cm]{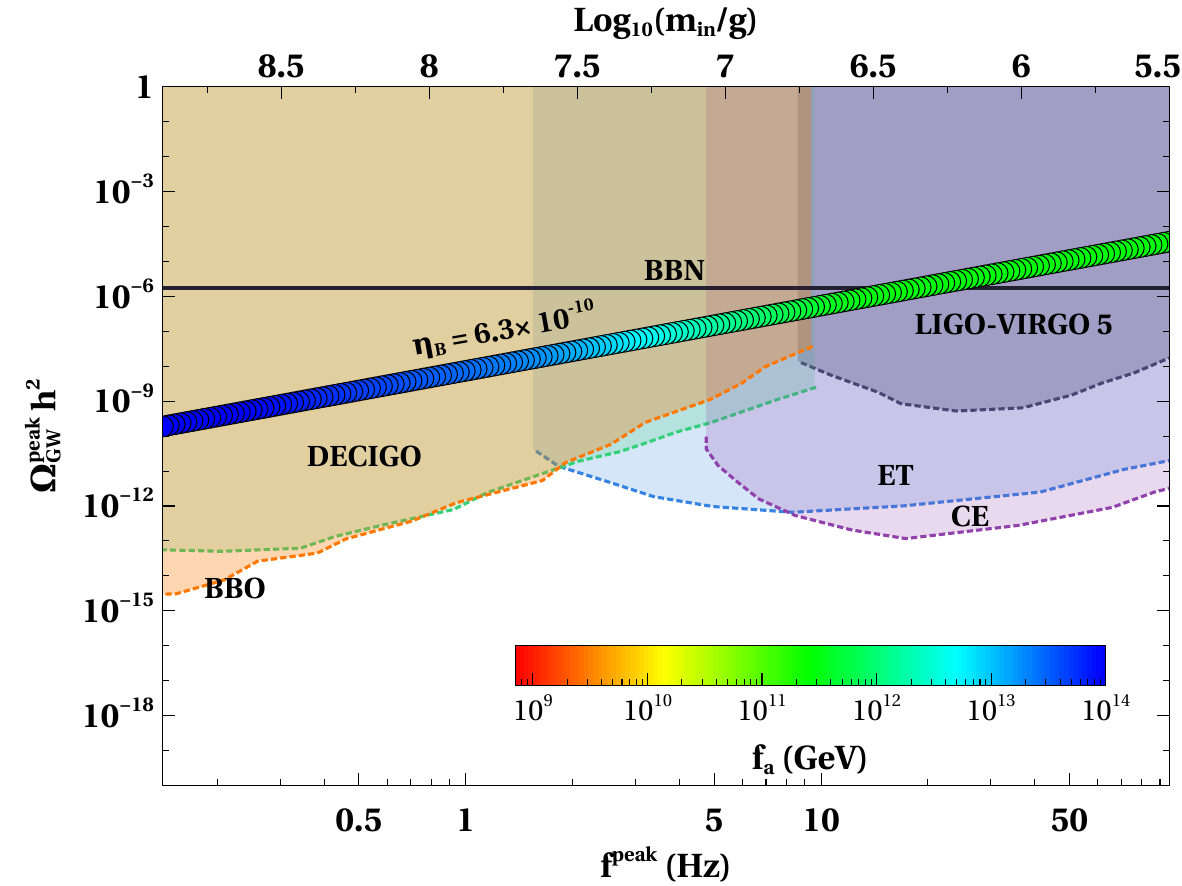}
    \caption{Left panel: $\Omega^{\rm peak}_{\rm GW} h^2$ versus $f^{\rm peak}$ for hierarchical scenario with three different $f_{a}$. Right panel: $\Omega^{\rm peak}_{\rm GW} h^2$ versus $f^{\rm peak}$ for the resonant scenario with maximal CP asymmetry for correct axion DM abundance.}
    \label{fig:GW_SPEC1}
\end{figure*}
Recently it has been pointed out that GWs arise from the inhomogeneities in the PBH distribution after they are formed  \cite{Papanikolaou:2020qtd, Domenech:2020ssp, Domenech:2021wkk, Domenech:2021ztg, Papanikolaou:2022chm}. This particular source of GW is generally independent of the PBH production mechanism. PBHs typically follow an inhomogeneous spatial distribution after formation obeying Poisson statistics \cite{Papanikolaou:2020qtd}. Such spatial inhomogeneities induce GW at second order once PBH dominate the energy density of the Universe \cite{Papanikolaou:2020qtd, Domenech:2020ssp}. The amplitude of the GW gets further enhanced during PBH evaporation \cite{Domenech:2020ssp, Inomata:2019ivs}. The  amplitude of such GW is given by \cite{Domenech:2020ssp, Domenech:2021wkk, Borah:2022iym, Barman:2022pdo}
\begin{equation}
    \Omega_{\rm gw}(t_0,f)\simeq \Omega_{\rm gw}^{\rm peak}\left(\frac{f}{f^{\rm peak}}\right)^{11/3}\Theta
\left(f^{\rm peak}-f\right),\label{eqn:omgw}
\end{equation}
where $ \Omega_{\rm gw}^{\rm peak}$ is the  peak amplitude which depends on the initial PBH mass and energy fraction as
\begin{equation}
    \Omega_{\rm gw}^{\rm peak}\simeq 2\times 10^{-6} \left(\frac{\beta}{10^{-8}}\right)^{16/3}\left(\frac{m_{\text{in}}}{10^7 \rm g}\right)^{34/9}.\label{eqn:omgpeak}
\end{equation}
On the other hand, the peak frequency $f^{\rm peak}$ which is related to the mean separation among PBH is given by 
\begin{equation}
    f^{\rm peak}\simeq 1.7\times 10^3\,{\rm Hz}\,\left(\frac{m_{\text{in}}}{10^4 \rm g}\right)^{-5/6}.\label{eqn:fpk}
\end{equation}

In order for the GW not to contradict the  limits on $N_{\rm eff}$ from BBN and CMB observations, the following constraint on $\beta$ must be satisfied
\begin{eqnarray}\label{beta_bound}
    \beta \lesssim 1.1\times 10^{-6} \left(\frac{m_{\rm in}}{10^4 \rm g}\right)^{-17/24}.
\end{eqnarray}
Note interestingly that Eq.\eqref{eq:betaM1}, \eqref{eq:beta_reso}, and \eqref{eqn:omgpeak} relate leptogenesis scale and $\Omega_{\rm gw}^{\rm peak}$ by trading $\beta$, leading to the relation
\begin{eqnarray}
    \Omega_{\rm gw}^{\rm peak}\simeq 6.29\times 10^{-22} \left(\frac{f_{a}}{10^{12} \rm{  GeV}}\right)^{16/3}\left(\frac{10^7 \rm g}{m_{\text{in}}}\right)^{14/9}\label{eqn:omgpeak_hie}
\end{eqnarray}
for hierarchical leptogenesis, and
\begin{eqnarray}
     \Omega_{\rm gw}^{\rm peak}\simeq 1.71\times 10^{-7} \left(\frac{\epsilon_1}{0.1}\right)^{16/3}\left(\frac{10^7 \rm g}{m_{\text{in}}}\right)^{14/9}\label{eqn:omgpeak_reso}
\end{eqnarray}
for resonant leptogensis, where for the latter scenario $\epsilon_1$ is expressed as a function of $f_a$ and mass splitting $\sim (M_2-M_1) M_2^{-1}$ between two degenerate RHNs. To relate GW with a successful cogenesis, Eq.\eqref{eqn:omgpeak_hie} and \eqref{eqn:omgpeak_reso} are required to be analysed with the constraint from correct axion DM production. Similar to the previous section, this has been implemented by solving the set of Boltzmann equations for axion and PBH (cf. Appendix\ref{appen1}).

In Fig.\ref{fig:GW_SPEC}, we show the GW spectra for the hierarchical (left panel) and the resonant scenario (right panel), along with the sensitivity of different GW experiments such as advanced LIGO-VIRGO/LIGO 5 \cite{LIGOScientific:2014pky}, LISA\,\cite{2017arXiv170200786A}, DECIGO \cite{Kawamura:2006up} and CE\,\cite{LIGOScientific:2016wof}. The spectra are plotted using Eq.\eqref{eqn:omgw} along with Eq.\eqref{eqn:omgpeak_hie} and \eqref{eqn:omgpeak_reso}. The corresponding $\{m_{\rm in}, f_{a}\}$ values are marked in the legend. Each spectrum corresponds to a different DM relic density but is consistent with the observed baryon asymmetry. Note that the signal strength is relatively weaker for the hierarchical leptogenesis case (left panel) when the model produces the correct DM relic (the magenta) curve. We however would like to mention that future planned detectors such as UDECIGO (UDECIGO-corr) \cite{Sato:2017dkf, Ishikawa:2020hlo} can test such signal.  On the other hand, for the resonant case, one obtains much stronger signal strength (right) without overproducing the DM. The reason being, as discussed in the previous section, resonant leptogenesis scenario requires a large value of $\beta$, and the peak amplitude goes as $ \Omega_{\rm gw}^{\rm peak}\sim \beta^{16/3}$, see Eq.\eqref{eqn:omgpeak}.

Plots in Fig.\ref{fig:GW_SPEC1}, represent an enlarged parameter space covering $\Omega^{\rm peak}_{\rm GW}h^{2}$ and $f^{\rm peak}$ for a wide range of $m_{\rm in}$. The left panel plot represents the hierarchical leptogenesis leading to successful BAU. The solid (dashed)  segment in the curves represents the correct (overproduced) DM abundance. As discussed earlier, because resonant cases allow strong amplitude GW, we find it sufficient to show the GW spectra only for correct DM abundance (right panel).


Note also that, although arguable \cite{Baeza-Ballesteros:2023say}, one can potentially have  GWs from axion-strings \cite{Gorghetto:2021fsn} in this model. Due to the PBH domination, our scenario can marginally access the values $f_a\gtrsim 10^{14}$ GeV to produce GWs from axion cosmic strings. However, we do not discuss this spectrum because, in the presence of PBH, a string-PBH network evolution requires numerical simulation \cite{Vilenkin:2018zol}. 

\subsection{Signatures on $  g_{a\gamma}-m_a$ plane} 
\label{sec6}
\begin{figure}
    \centering
        \includegraphics[width=7 cm, height=6cm]{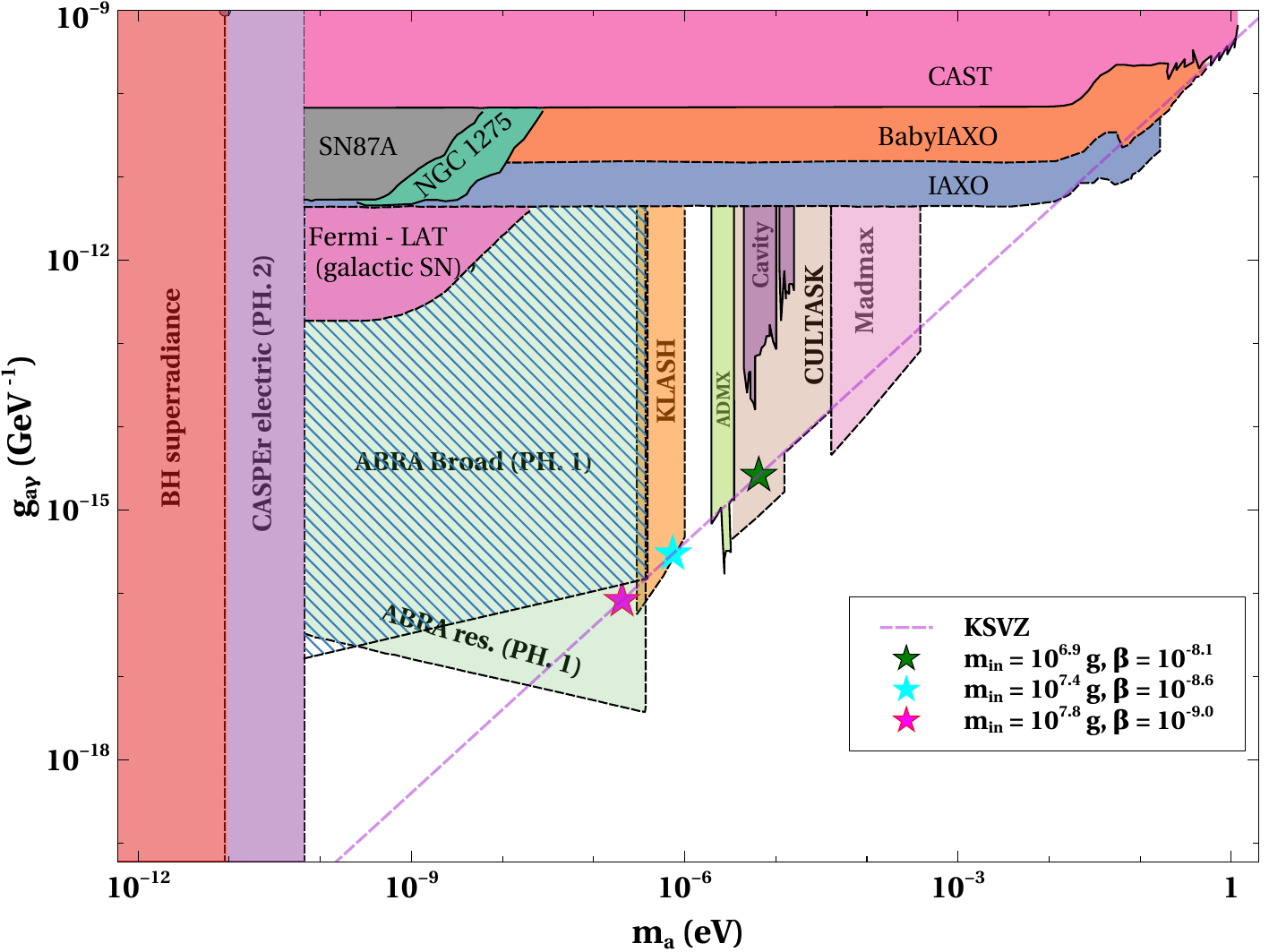} 
            \includegraphics[width=7 cm, height=6cm]{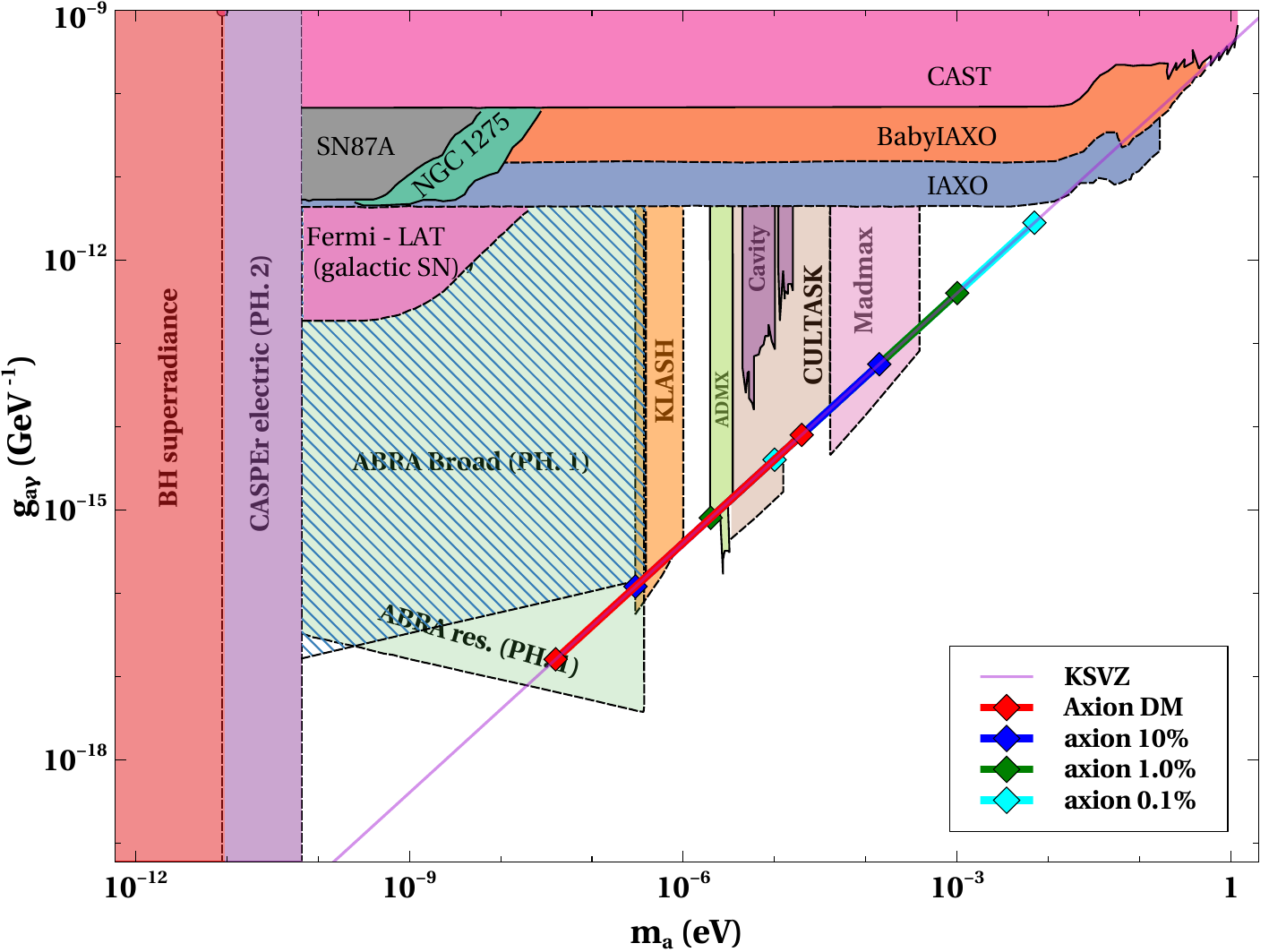}
    \caption{Left panel: Axion-photon coupling versus axion mass in presence of PBH for three different sets of \{$m_{\rm in}, \beta$ \} denoted by green, cyan, and magenta colored star-shaped points. Cogenesis with resonant leptogenesis is possible for all three points. Also, all three sets of \{$m_{\rm in},\beta$\} come within future GW experiments (see related text for more details). Right panel: Constraints on axion-photon coupling in the presence of PBH for four different axion contributions to DM.}
    \label{fig:photon_coupling_without_pbh}
\end{figure}

The axion detection prospects in most experiments primarily rely on axion-photon coupling. The axion-photon coupling leading to $\frac{1}{4} g_{a\gamma} a F\Tilde{F}$ is given by (same as KSVZ model) \cite{Srednicki:1985xd}
\begin{eqnarray}
     g_{a\gamma} = -\frac{\alpha}{2 \pi f_{a}} \left(\frac{2}{3} \frac{4 m_{d}+m_{u}}{m_{u} + m_{d}}\right) = -1.92 \frac{\alpha}{2 \pi f_{a}},
 \end{eqnarray}
with $\alpha$ being the fine-structure constant. Fig.\ref{fig:photon_coupling_without_pbh} shows the axion-photon coupling with axion mass including the bounds and sensitivities of different experiments. The current experimental bound on the axion-photon coupling from various experiments or observables are shown by the solid color lines (CAST \cite{CAST:2007jps, CAST:2017uph}, SN87A \cite{PhysRevLett.60.1797, PhysRevD.39.1020, PhysRevLett.60.1793}, NGC 1275 \cite{Fermi-LAT:2016nkz}, ADMX \cite{ADMX:2006kgb, Stern:2016bbw, ADMX:2019uok}) whereas future experimental sensitivies are shown by the dashed lines (CASPEr \cite{Budker:2013hfa}, KLASH \cite{Alesini:2017ifp,Alesini:2019nzq, Alesini:2023qed}, ABRACADABRA \cite{Kahn:2016aff, Ouellet:2018beu}, CULTASK \cite{Lee:2020cfj, Semertzidis:2019gkj}, MADMAX \cite{Caldwell:2016dcw}, IAXO \cite{Vogel:2013bta, IAXO:2019mpb},  Fermi-LAT \cite{Meyer:2016wrm}, BH superradiance \cite{Cardoso:2018tly}).

 The left panel of Fig.\ref{fig:photon_coupling_without_pbh} shows the constraints on $g_{a\gamma}$ in the presence of PBH for three different axion mass associated with three different sets of \{$m_{\rm in}, \beta$\}. These three points are indicated by green, cyan, and magenta colored star-shaped point ($\bigstar$).  For all three points, both DM relic and correct baryon asymmetry (with resonant leptogenesis) are satisfied. While the green point comes within the sensitivity of future CULTASK experiment, the cyan and magenta points can be probed by future KLASH and ABRACADABRA experiments. Apart from their signatures in axion detection experiments, the three sets of \{$m_{\rm in},\beta$\} also lie within the sensitivity of future  DECIGO. This provides a complementary probe of cogenesis in terms of GWs and axion detection experiments. Note that for a smaller value of $\beta$ (cf. left panel of  Fig.\ref{fig:mbh_vs_beta}), the scenario of hierarchical leptogenesis can also be within the sensitivity of axion experiments, however with a weaker GW signal.
 
 We conclude with the following remark: Because we consider axions to constitute the entire DM density, the overall scenario is highly constrained. This framework, however, can naturally allow one of the RHNs to make up the dominant dark matter density while keeping axion abundance subdominant. In which case, the predictions of this scenario would be very different. Interested readers may look up to a detailed discussion presented on this in Appendix \ref{appb}, \ref{appd}, and \ref{appc}.  Here for example, in the right panel of Fig.\ref{fig:photon_coupling_without_pbh}, we show the constraint for four different situations where axion contribution to DM is $100\%$, $10\%$, $1\%$ and $0.1\%$ denoted by red, blue, green, and cyan line respectively. The minimum $f_{a}$ is obtained for PBH mass with $T_{\rm ev}>T_{\rm osc}$ ( i.e. $m_{\rm in} \lesssim 10^{6.5}$g ). The maximum $f_{a}$ is obtained for PBH mass with $T_{\rm ev} \sim T_{\rm BBN}$ and for maximum allowed $\beta$ from Eq.\eqref{beta_bound} (i.e. $m_{\rm in} \sim 10^{8.6}$ g and $\beta \sim 6\times 10^{-10}$). While in this scenario the vanilla leptogenesis parameter space is still accessible, a striking difference would be to have only resonant leptogenesis for successful baryogenesis.

\section{Conclusion}
\label{sec6}
The presence of an early matter-dominated epoch in the early Universe allows a larger axion decay constant $f_a$ consistent with correct DM abundance. We implement this idea in a SMASH-like setup where the $f_a$ and seesaw or leptogenesis scale are equivalent. We show that contrary to the EMD provided, e.g., by a long-lived field, a PBH-domination scenario is markedly testable with gravitational waves alongside the axion experiments that intend to detect low-mass axions ($m_a\lesssim 10^{-5}$ eV). In particular, this setup opens up a novel opportunity to test unflavored leptogenesis scales $f_a\gtrsim 10^{12}$ GeV, which are otherwise neither testable from measurements of neutrino mixing parameters like Dirac CP phase nor accessible directly at terrestrial experiments. A hierarchical (resonant) leptogenesis scenario produces weaker (stronger) gravitational waves within the frequency band of planned GW detectors. A successful cogenesis renders extremely constrained model parameter space and thereby robust predictions. In addition to GW and axion detection prospects, we can also have observable dark radiation in future CMB experiments due to the axions produced from PBH evaporation. This keeps our cogenesis setup verifiable at gravitational wave, axion detectors as well as CMB experiments.

\section*{Acknowledgements}
We thank Satyabrata Datta for the discussion at the initial stages of this work. The work of DB is supported by the Science and Engineering Research Board (SERB), Government of India grants MTR/2022/000575 and CRG/2022/000603. The work of ND is supported by the Ministry of Education, Government of India via the Prime Minister's Research Fellowship (PMRF) December 2021 scheme. ND would like to thank Disha Bandyopadhyay for useful discussions. The work of SJD was supported by IBS under the project code, IBS-R018-D1. SJD would also like to thank the support and hospitality of the Tata Institute of Fundamental Research (TIFR), Mumbai, where a part of this project was carried out.

\appendix
\section{Axion field evolution}
\label{appen1}

PBH, after formation, loses its energy via Hawking evaporation. The combined Boltzmann equations for the system can be written as
\begin{eqnarray}
\frac{d \rho_{\rm in}}{dt} + 3 \mathcal{H} \rho_{\rm BH} = \frac{1}{m_{\rm BH}} \frac{d m_{\rm BH}}{dt} \rho_{\rm BH} \label{eqn:beq}\\
\frac{d \rho_{\rm r}}{dt} + 4 \mathcal{H} \rho_{r} = - \frac{1}{m_{in}} \frac{d m_{\rm BH}}{dt} \rho_{\rm BH}.
\end{eqnarray}
The PBH mass loss rate is given by
\begin{eqnarray}
    \frac{1}{m_{\rm BH}} \frac{d m_{\rm BH}}{dt} = - \frac{\mathcal{G} g_{H}(T_{H})}{30720 \pi G^2 m^3_{\rm BH}}.
\end{eqnarray}
Due to PBH evaporation, entropy injection takes place, leading to non-conservation of entropy. The entropy injection can be tracked via 
\begin{eqnarray}
    \frac{ds}{dt} + 3 \mathcal{H} s = - \frac{1}{m_{\rm BH}} \frac{d m_{\rm BH}}{dt} \frac{\rho_{\rm BH}}{T}.
\end{eqnarray}
The evolution of axion field $a$ in the early Universe can be written as
\begin{eqnarray}
    \Ddot{a} + 3\mathcal{H} \Dot{a} + \frac{1}{R^2(t)} \nabla ^2 a + \frac{\partial V(a, T)}{\partial a} = 0,
\end{eqnarray}
where 
\begin{eqnarray}
    V(a, T) = f^2_{a} m^2_{a}(T) \left(1-\cos{\left(\frac{a}{f_{a}}\right)}\right)
\end{eqnarray}
and $R$ denotes the scale factor. The initial axion angle, $\theta = \frac{a}{f_{a}}$ is frozen in until the oscillation temperature of axion which can be estimated by comparing Hubble expansion rate to the temperature dependent axion mass $\mathcal{H} (T_{\rm osc}) = m_{a} (T_{\rm osc})$. To get the contribution of axion to the current energy budget, we solve the above equations following the procedure in the reference \cite{Mazde:2022sdx}. The procedure is to first estimate the oscillation temperature $T_{\rm osc}$ in presence of PBH. With this, number density of axion at $T_{\rm osc}$ can be calculated following
\begin{eqnarray}
    n_{a}(T_{\rm osc}) = \frac{\rho_{a}(T_{\rm osc})}{m_{a}(T_{\rm osc})} = \frac{1}{{m_{a}(T_{\rm osc})}} \left(\frac{1}{2}\Dot{a}^2 + \frac{1}{2} |\nabla a|^2 + V(a)\right).
\end{eqnarray}
Due to entropy dilution, the equation for $n_{a}$ becomes 
\begin{eqnarray}
    \frac{1}{n_{a}} \frac{d(n_{a}/s)}{dt} = \frac{1}{m_{\rm BH}} \frac{d m_{\rm BH}}{dt} \frac{\rho_{\rm BH}}{T s^2}.
\end{eqnarray}
The above equation is solved from $T \gtrsim T_{\rm osc}$ to a sufficiently small temperature when PBH completely evaporates. 
\section{Multi-component Dark Matter}\label{appb}
\begin{figure}
    \includegraphics[scale=0.42]{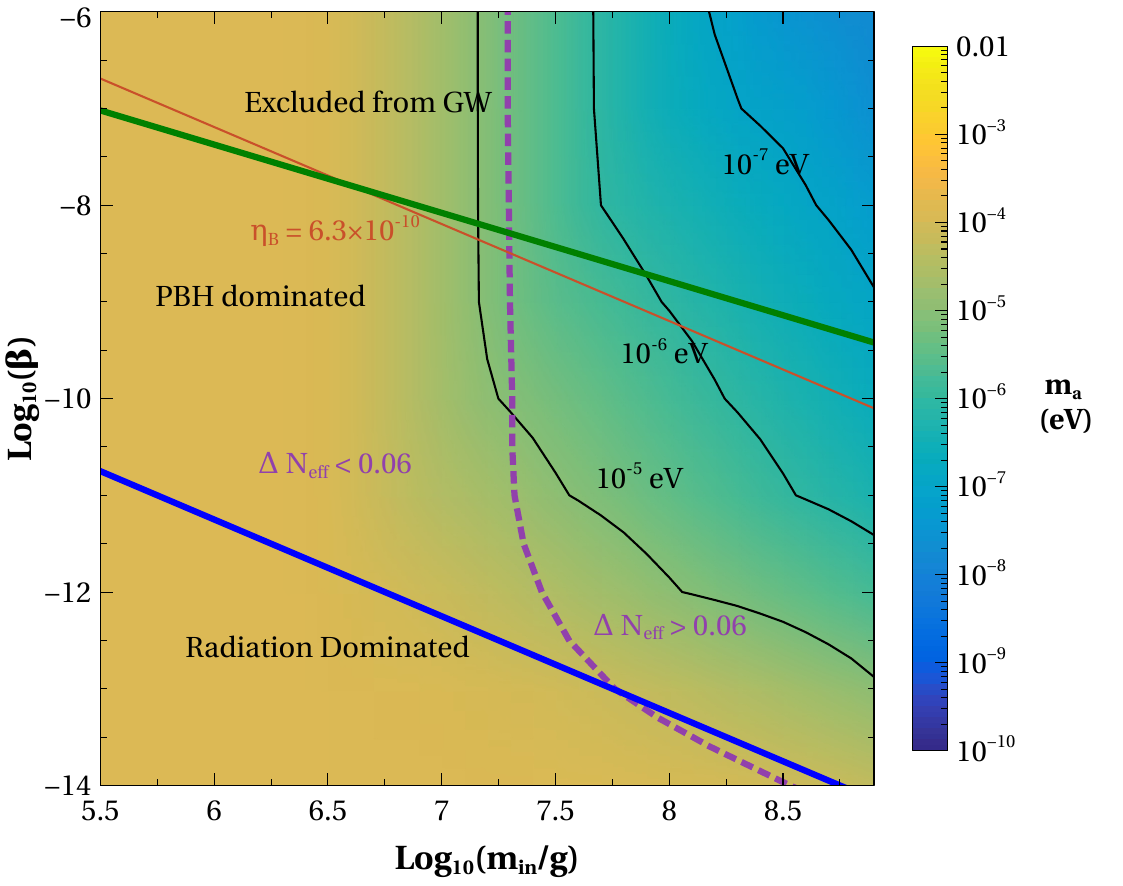}
            \includegraphics[scale=0.42]{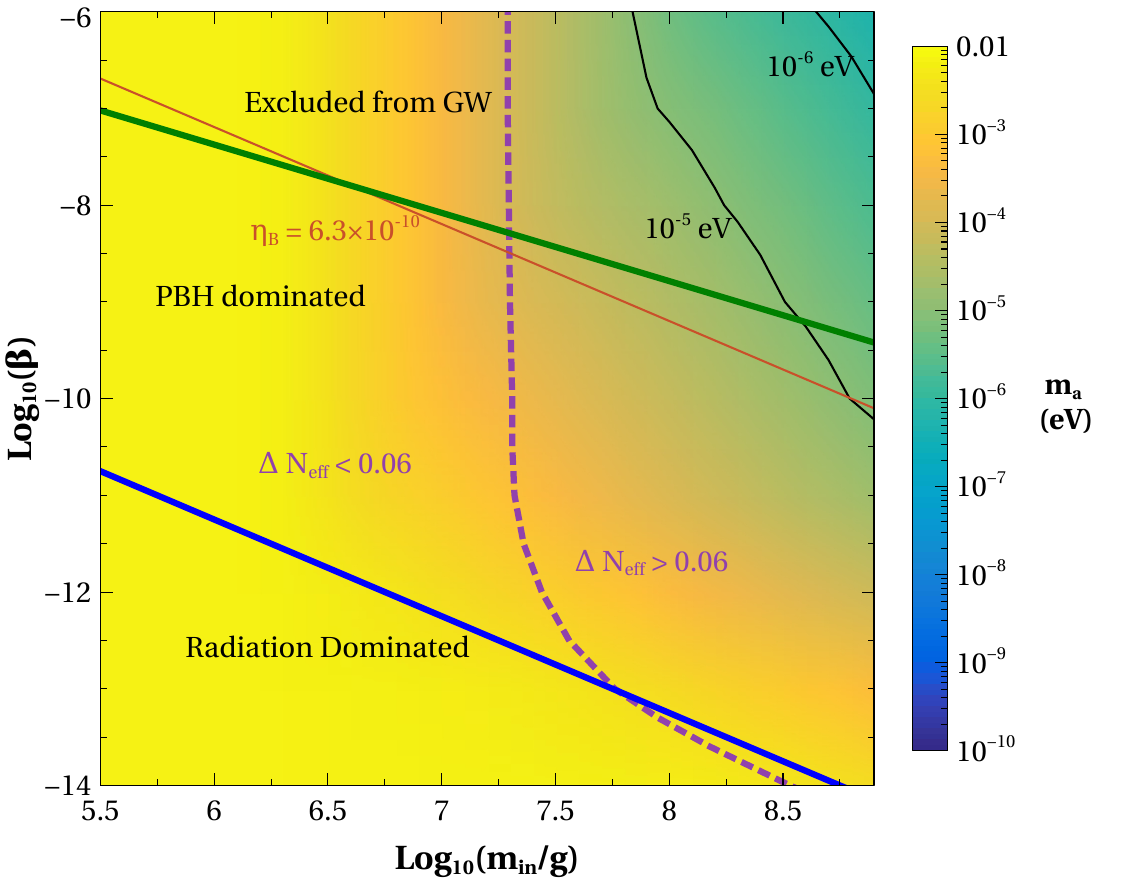}
            \caption{Constraint on $m_{\rm in}$ versus $\beta$ plane considering resonant leptogenesis. Axion constitutes $10\%, 0.1\%$ of total DM abundance in left and right panels respectively. The red solid lines in the plots denote the parameter space that is consistent with both DM relic and baryon asymmetry. The region excluded from GW is separated by solid green line. The magenta dashed line separate the regions that can be probed by future CMB-S4 experiment due to contribution of hot axion on $N_{\rm eff}$.}
    \label{fig:mbh_vs_beta_multi_10}
\end{figure}
If axion does not constitute the entire DM, the allowed parameter space discussed so far can change significantly. The deficit in total DM abundance can be filled up by one of the RHNs, as discussed briefly in Appendix \ref{appd}. From the left panel of Fig.\ref{fig:mbh_vs_beta}, for a particular value of $m_{\rm in}$ and $\beta$, one needs a larger value of axion mass (correspondingly smaller value of $f_{a}$) to have sub-dominant axion DM. However, at the same time, a smaller value of $f_{a}$ also reduces the CP asymmetry parameter according to Eq.\eqref{eq:epsilon}. As a result, baryon asymmetry also reduces. Thus, for a particular PBH mass, one needs to consider a smaller value $\beta$ so that dilution is also less to obtain correct baryon asymmetry. However, it turns out that for this particular scenario, even if one reduces $\beta$ to $\beta_{c}$ (where $\beta_{c}$ distinguishes PBH-dominated and radiation-dominated region), the correct baryon asymmetry can not be obtained. The maximum baryon asymmetry that can be obtained is $\sim 4\times 10^{-10}$ which is less than the observed asymmetry, viz., $6.3\times 10^{-10}$. All the region above the $\beta = \beta_{c}$ line gives $\eta_{\rm B} < 4\times10^{-10}$. For the sub-dominant axion DM scenario, one therefore must consider the resonant leptogenesis approach to generate the correct BAU. In Fig.\ref{fig:mbh_vs_beta_multi_10} we show the results for resonant leptogenesis by considering the percentage of axion DM as a free parameter. The solid red line in these plots correspond to the parameter space consistent with observed baryon asymmetry for $\epsilon_{1} = 0.1$.

Note that, in the multi-component DM scenario, because the axion DM is a subdominant component, we can have smaller $f_a$ values leading to axion mass range in the solar axion ballpark 
 \cite{IAXO:2019mpb,TASTE:2017pdv}, making this scenario pretty different from the case of axion constituting $100\%$ DM. As an aside, let us mention that there have been efforts to obtain the entire dark matter relic only in the form of axions for a smaller value of decay constant ($f_a\sim 10^{8}$GeV) via parametric resonance production of axions. see. e.g., Refs.\cite{Co:2017mop,Co:2020dya,Ramazanov:2022kbd}.


\section{Production of Right-Handed Neutrino Dark Matter}\label{appd}
As two RHNs are sufficient to generate neutrino mass and BAU via leptogenesis, the other RHN, if sufficiently long-lived, can constitute some part of the total DM abundance. Due to the presence of PBH and RHN coupling to axion as well as SM leptons, we can have three\footnote{It is also possible to generate light RHN DM via active-sterile neutrino oscillation \cite{Dodelson:1993je}, though it faces severe X-ray constraints. See \cite{Drewes:2016upu} for a review.} different production channels for RHN DM namely, (i) gravitational production from PBH evaporation \cite{Gondolo:2020uqv, Bernal:2020bjf, Green:1999yh, Khlopov:2004tn, Dai:2009hx, Allahverdi:2017sks, Lennon:2017tqq, Hooper:2019gtx, Sandick:2021gew, Fujita:2014hha, Datta:2020bht, JyotiDas:2021shi,Barman:2021ost, Barman:2022gjo, Barman:2022pdo, Cheek:2021odj, Chaudhuri:2023aiv, Borah:2024lml}, (ii) production via axion portal interactions \cite{Ghosh:2023tyz, Bharucha:2022lty, Fitzpatrick:2023xks}, (iii) freeze-in production via Yukawa portal interactions \cite{Borah:2020wyc, Datta:2021elq}.

For the axion decay constant we have in our setup, superheavy RHN DM is more natural which is produced dominantly via PBH evaporation. For the case when $M_{\rm DM}< T^{\rm in}_{\rm BH}$, correct relic abundance needs DM mass in the range of $\sim$ GeV. However, for this mass range of DM produced by PBH evaporation, the DM turns out to be hot and hence disfavored from structure formation constraints. For the other limiting case $M_{\rm DM}> T^{\rm in}_{\rm BH}$, the DM mass is greater than $10^{9}$ GeV, keeping it in the superheavy ballpark, a natural consequence of the high PQ symmetry breaking scale

\begin{figure*}
    \includegraphics[width=7 cm, height=6.7cm]{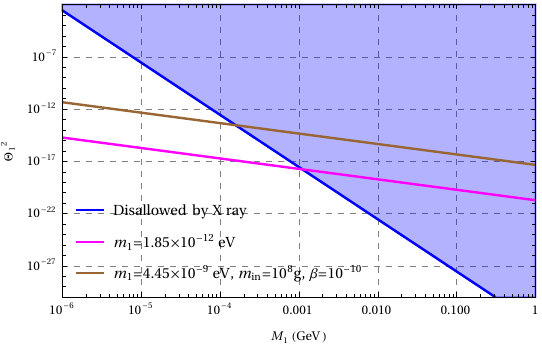}
    \includegraphics[scale=0.45]{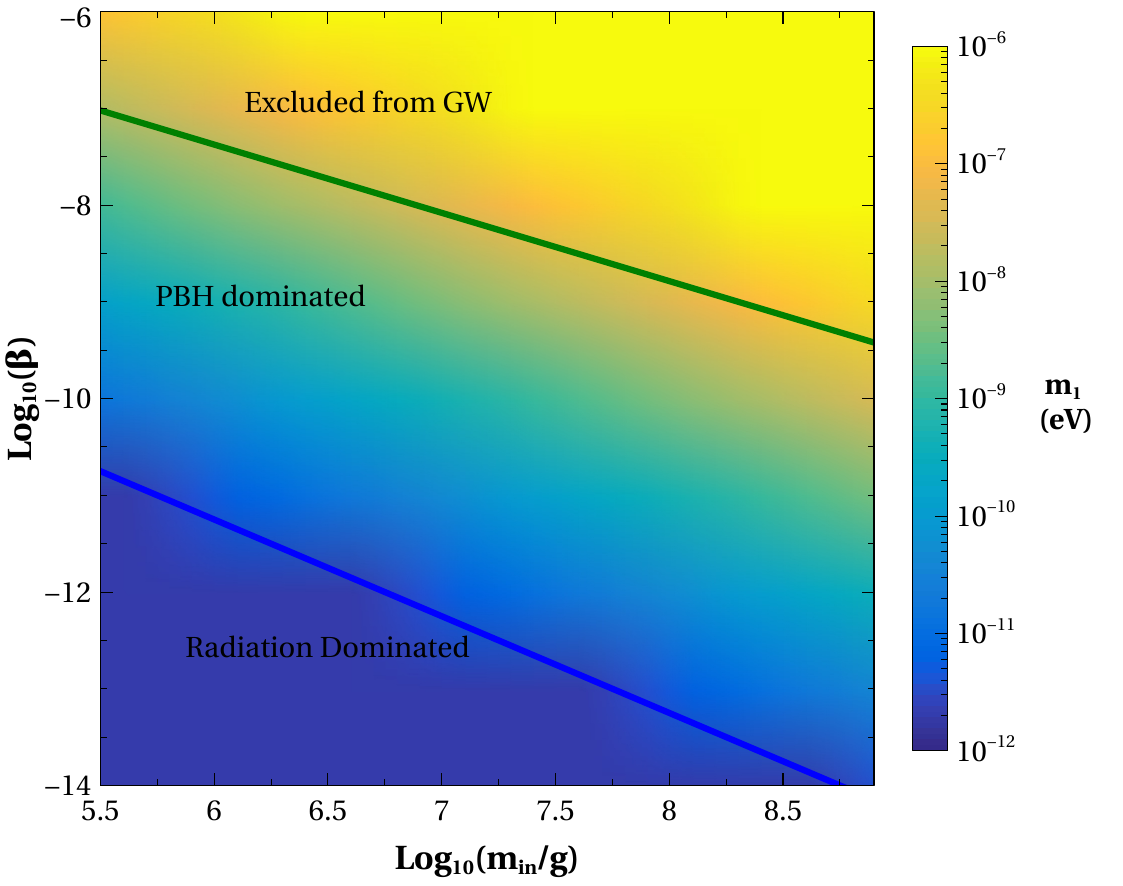}
    \caption{Left panel: $\theta^2_{1}$ versus $M_{1}$ for two different scenarios: without PBH and with PBH ($m_{\rm in}=10^{8}$ g and $\beta=10^{-10}$)  shown by magenta and brown color respectively. The shaded blue color region represents the excluded region by X-ray experiments. Right panel: Lightest neutrino mass $m_{1}$ in PBH parameter space $\beta - m_{\rm in}$ with $N_{1}$ abundance $\Omega_{N_{1}}h^2=0.12$.}
    \label{fig:RHN_DM}
\end{figure*}

However, if we allow fine-tuning of RHN DM mass, keeping it much lower than the PQ scale, other production mechanisms like axion portal can dominate. Since, axion portal coupling of RHN will be suppressed for allowed mass range of thermal DM below unitarity limit \cite{Griest:1989wd}, such production is likely to occur via freeze-in mechanism from the SM bath. For $f_{a} \sim 10^{12}$ GeV and reheat temperature $T_{\rm RH} \sim 10^{12}$ GeV, we require DM mass to be around $10^{5}$ GeV for observed DM relic to be produced via UV freeze-in. For heavier DM mass, it is likely to be overproduced via axion portal interactions as long as $T_{\rm RH} > M_{\rm DM}$. On the other hand, RHN DM of mass $\sim 10^{5}$ GeV will also be overproduced from PBH evaporation as mentioned above. This pushes RHN DM to low mass where it can be produced efficiently via Yukawa portal interactions while gravitational and axion portal contributions remain sub-dominant. In this regime, correct relic can be achieved via IR freeze-in production of $N_{1}$ from the SM bath. As shown in earlier works \cite{Borah:2020wyc, Datta:2021elq}, such freeze-in production is dominated by $W^{\pm}, Z$ and SM Higgs decays facilitated via active-sterile neutrino mixing $\theta_1$. The mixing angle can be tuned by choosing the lightest active neutrino mass $m_1$. The same active-sterile mixing also leads to decay of $N_1$ into SM particles. Although the abundance of $N_{1}$ only depends on $m_{1}$, the decay rate of $N_{1}$ depends on both $m_{1}$ and $M_{1}$. This gives a constraint on the $N_{1}$ mass, as shown in the left panel of Fig.\ref{fig:RHN_DM}. The magenta-colored line represents the situation where PBH do not dominate the energy density of the Universe (i.e. no additional entropy dilution). The brown colored line denotes the situation where entropy dilution is present from PBH evaporation. Due to the presence of entropy dilution, $N_{1}$ should be produced over-abundantly requiring a larger value of active-sterile mixing angle. As the decay of $N_{1}$ is also dependent on $\theta^{2}_{1}$, so $M_{1}$ is more tightly constraint in the presence of entropy dilution as seen from the left plot of Fig.\ref{fig:RHN_DM}. Here both the magenta and the brown color lines give the correct relic density. In the right panel plot, $m_{1}$ is shown in the $\beta - m_{\rm in}$ plane for $\Omega_{N_{1}} h^2 =0.12$, correlating neutrino mass with PBH parameters from correct DM relic criteria.

\begin{figure}
    \centering
    \includegraphics[scale=0.45]{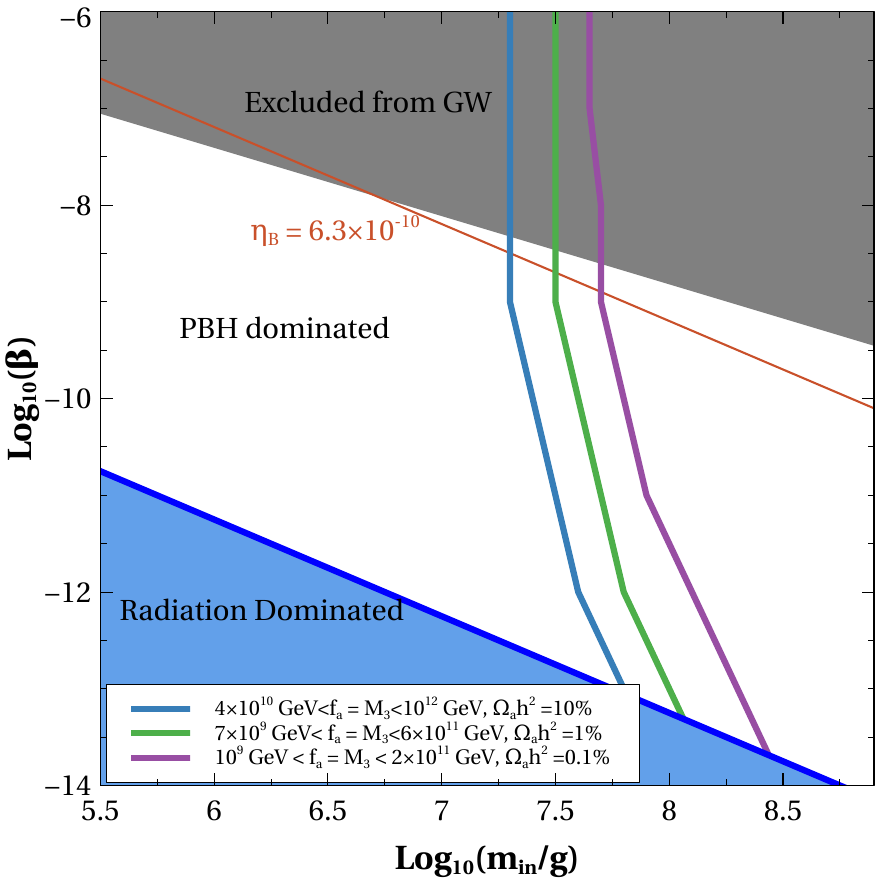}
    \caption{For $M_{3} \gg T_{\rm RH} > M_{2} \sim M_{3}$. Blue, green, and magenta colored lines denote the different situations where $\Omega_{\rm a} h^2=10\%, \Omega_{\rm a} h^2=1\%$ and $\Omega_{\rm a} h^2=0.1\%$ respectively and $f_{a} = M_{3}$. The RHN $N_{3}$ constitutes the rest of DM abundance. The brown-colored solid line denotes the parameter space consistent with the baryon asymmetry for maximal CP asymmetry.}
    \label{fig:M3_fa}
\end{figure}

{\bf RHN DM with $T_{\rm RH} < M_{\rm DM}$:} 
As noticed in the above discussions, the RHN DM parameter space produced from PBH evaporation and axion portal interactions is anti-correlated, forcing the RHN DM mass to sub-GeV ballpark where it can be efficiently produced from Yukawa portal freeze-in mechanisms. However, this requires severe fine-tuning of RHN coupling with the PQ scalar such that $ M_1 \equiv M_{\rm DM} \ll f_a$. Such unnatural fine-tuning can, however, be avoided if we consider the heaviest RHN $N_{3}$ to be DM with mass $M_{3}=f_{a} > T_{\rm RH}$ which can not be produced from the bath. Leptogenesis is obtained from $N_{1}$ and $N_{2}$ with $M_{1} \simeq M_{2}$. For thermal leptogenesis, we assume $M_{1}\simeq M_{2} < T_{\rm RH} \ll M_{3}=f_{a}$. Although the PQ symmetry gets broken after e.g., slow-roll inflation, a non-instantaneous reheating phase \cite{Giudice:2000ex} can lead to a reheat temperature $T_{\rm RH} \ll f_a$. This leads to purely gravitational production of RHN DM from PBH evaporation. The DM mass produced from PBH evaporation is given by 
\begin{equation}
    M_{\rm DM} \simeq 4.5\times10^{3}  \left(\frac{0.12}{\Omega_{N_{3}}h^2}\right) \left(\frac{m_{\rm in}}{M_{\rm Pl}}\right)^{-5/2} \frac{M^{2}_{\rm Pl}}{\text{GeV}}.\label{eq:N3dm}
\end{equation}
Considering $M_{\rm DM}=f_{a}$, this gives a line in the $\beta - m_{\rm in}$ plane. In Fig.\ref{fig:M3_fa}, blue, green, and magenta colored lines correspond to the situations where $\Omega_{a}h^2=10\%, 1\%$ and $0.1\%$ of total DM abundance respectively. Note that $M_3 = M_{\rm DM}=f_a$ changes along these contours, following Eq.\eqref{eq:N3dm}.  The rest of the DM abundance is constituted by $N_{3}$. The observed BAU can be generated from resonant leptogenesis as discussed earlier. The parameter space satisfying the observed BAU for maximal CP asymmetry $\epsilon_{1} =0.1$ is indicated by the brown-colored solid line in Fig.\ref{fig:M3_fa}.

\section{GW prospects in multi-component DM scenario}\label{appc}
\begin{figure*}
    
    \centering
    \includegraphics[scale=0.37]{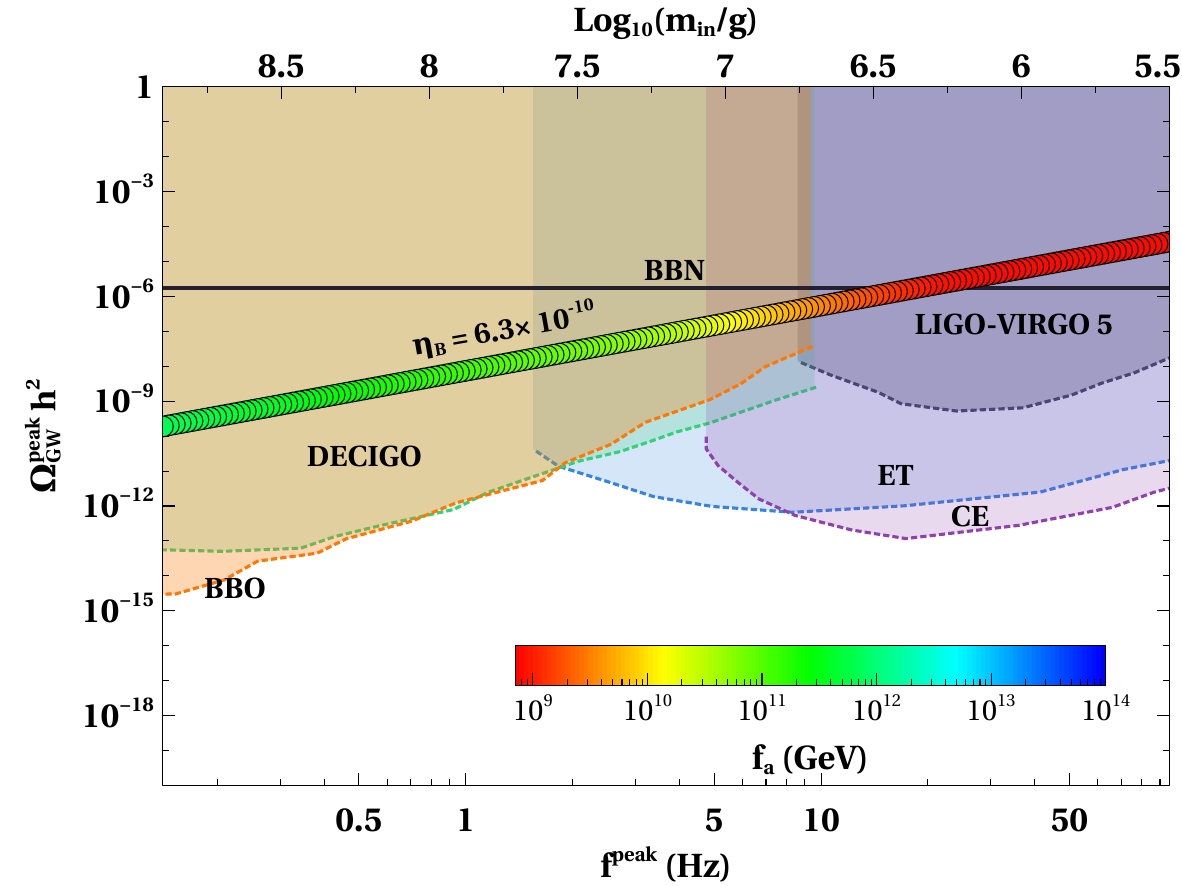}
    \includegraphics[scale=0.47]{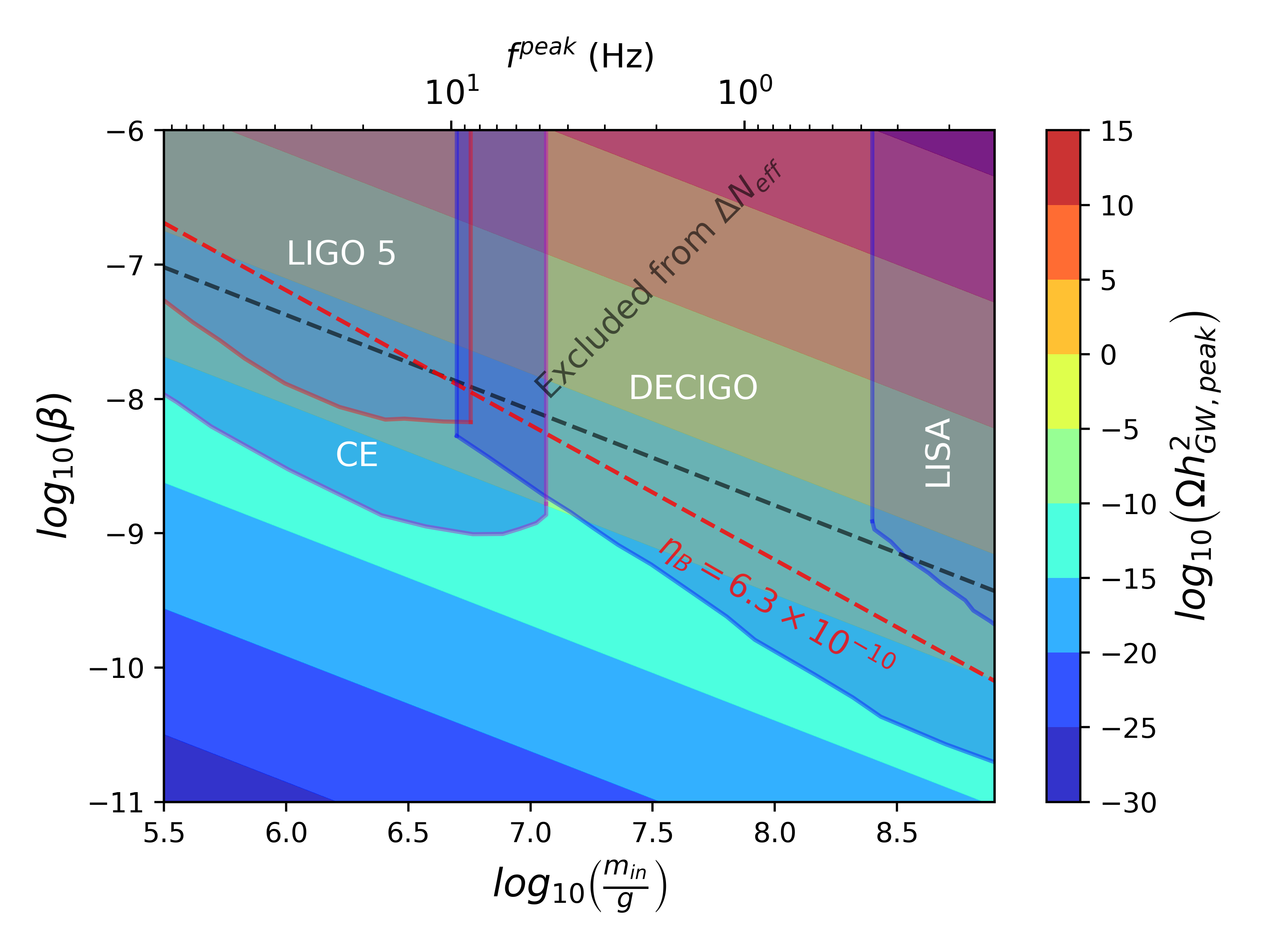}
    \caption{Left panel: $\Omega^{\rm peak}_{\rm GW} h^2$ versus $f^{\rm peak}$ for $0.1\%$ relative abundance of axion DM. Right Panel : Peak Gravitational Waves amplitude as a function of $m_{\rm in}$ and $\beta$ with different experimental sensitivities. The red dashed line denotes the correct baryon asymmetry value for resonant leptogenesis with maximal CP asymmetry.}
    \label{fig:GW}
\end{figure*}

Similar to the right panel of Fig.\ref{fig:GW_SPEC1}, we show the $\Omega^{\rm peak}_{\rm GW}h^{2}$ and $f^{\rm peak}$ for resonant leptogenesis scenario in the left panel of Fig.\ref{fig:GW} considering $0.1\%$ relative abundance of axion DM. In the right panel, we show another representation of the resonant scenario where the peak GWs amplitude are shown as a function of PBH mass and $\beta$ along with different GWs sensitivities. As expected, the red dashed colored line denoting correct leptogenesis for resonant scenario lies within the future GWs sensitivities. 

\bibliographystyle{JHEP}
\bibliography{ref, ref1, ref2, ref3}

\providecommand{\href}[2]{#2}\begingroup\raggedright\begin{thebibliography}{100}

\bibitem{Abel:2020pzs}
C.~Abel et~al., {\it {Measurement of the Permanent Electric Dipole Moment of
  the Neutron}},  {\em Phys. Rev. Lett.} {\bf 124} (2020), no.~8 081803,
  [\href{http://arxiv.org/abs/2001.11966}{{\tt arXiv:2001.11966}}].

\bibitem{Peccei:1977hh}
R.~D. Peccei and H.~R. Quinn, {\it {CP Conservation in the Presence of
  Instantons}},  {\em Phys. Rev. Lett.} {\bf 38} (1977) 1440--1443.

\bibitem{Peccei:1977ur}
R.~D. Peccei and H.~R. Quinn, {\it {Constraints Imposed by CP Conservation in
  the Presence of Instantons}},  {\em Phys. Rev. D} {\bf 16} (1977) 1791--1797.

\bibitem{Wilczek:1977pj}
F.~Wilczek, {\it {Problem of Strong $P$ and $T$ Invariance in the Presence of
  Instantons}},  {\em Phys. Rev. Lett.} {\bf 40} (1978) 279--282.

\bibitem{Weinberg:1977ma}
S.~Weinberg, {\it {A New Light Boson?}},  {\em Phys. Rev. Lett.} {\bf 40}
  (1978) 223--226.

\bibitem{Preskill:1982cy}
J.~Preskill, M.~B. Wise, and F.~Wilczek, {\it {Cosmology of the Invisible
  Axion}},  {\em Phys. Lett. B} {\bf 120} (1983) 127--132.

\bibitem{Abbott:1982af}
L.~F. Abbott and P.~Sikivie, {\it {A Cosmological Bound on the Invisible
  Axion}},  {\em Phys. Lett. B} {\bf 120} (1983) 133--136.

\bibitem{Dine:1982ah}
M.~Dine and W.~Fischler, {\it {The Not So Harmless Axion}},  {\em Phys. Lett.
  B} {\bf 120} (1983) 137--141.

\bibitem{Raffelt:2006cw}
G.~G. Raffelt, {\it {Astrophysical axion bounds}},  {\em Lect. Notes Phys.}
  {\bf 741} (2008) 51--71, [\href{http://arxiv.org/abs/hep-ph/0611350}{{\tt
  hep-ph/0611350}}].

\bibitem{Caputo:2024oqc}
A.~Caputo and G.~Raffelt, {\it {Astrophysical Axion Bounds: The 2024 Edition}},
   in {\em {1st Training School of the COST Action COSMIC WISPers (CA21106)}},
  1, 2024.
\newblock \href{http://arxiv.org/abs/2401.13728}{{\tt arXiv:2401.13728}}.

\bibitem{Dine:1981rt}
M.~Dine, W.~Fischler, and M.~Srednicki, {\it {A Simple Solution to the Strong
  CP Problem with a Harmless Axion}},  {\em Phys. Lett. B} {\bf 104} (1981)
  199--202.

\bibitem{Zhitnitsky:1980tq}
A.~R. Zhitnitsky, {\it {On Possible Suppression of the Axion Hadron
  Interactions. (In Russian)}},  {\em Sov. J. Nucl. Phys.} {\bf 31} (1980) 260.

\bibitem{Kim:1979if}
J.~E. Kim, {\it {Weak Interaction Singlet and Strong CP Invariance}},  {\em
  Phys. Rev. Lett.} {\bf 43} (1979) 103.

\bibitem{Shifman:1979if}
M.~A. Shifman, A.~I. Vainshtein, and V.~I. Zakharov, {\it {Can Confinement
  Ensure Natural CP Invariance of Strong Interactions?}},  {\em Nucl. Phys. B}
  {\bf 166} (1980) 493--506.

\bibitem{Kawasaki:2013ae}
M.~Kawasaki and K.~Nakayama, {\it {Axions: Theory and Cosmological Role}},
  {\em Ann. Rev. Nucl. Part. Sci.} {\bf 63} (2013) 69--95,
  [\href{http://arxiv.org/abs/1301.1123}{{\tt arXiv:1301.1123}}].

\bibitem{Fukugita:1986hr}
M.~Fukugita and T.~Yanagida, {\it {Baryogenesis Without Grand Unification}},
  {\em Phys. Lett. B} {\bf 174} (1986) 45--47.

\bibitem{Servant:2014bla}
G.~Servant, {\it {Baryogenesis from Strong $CP$ Violation and the QCD Axion}},
  {\em Phys. Rev. Lett.} {\bf 113} (2014), no.~17 171803,
  [\href{http://arxiv.org/abs/1407.0030}{{\tt arXiv:1407.0030}}].

\bibitem{Ipek:2018lhm}
S.~Ipek and T.~M.~P. Tait, {\it {Early Cosmological Period of QCD
  Confinement}},  {\em Phys. Rev. Lett.} {\bf 122} (2019), no.~11 112001,
  [\href{http://arxiv.org/abs/1811.00559}{{\tt arXiv:1811.00559}}].

\bibitem{Croon:2019ugf}
D.~Croon, J.~N. Howard, S.~Ipek, and T.~M.~P. Tait, {\it {QCD baryogenesis}},
  {\em Phys. Rev. D} {\bf 101} (2020), no.~5 055042,
  [\href{http://arxiv.org/abs/1911.01432}{{\tt arXiv:1911.01432}}].

\bibitem{Co:2019wyp}
R.~T. Co and K.~Harigaya, {\it {Axiogenesis}},  {\em Phys. Rev. Lett.} {\bf
  124} (2020), no.~11 111602, [\href{http://arxiv.org/abs/1910.02080}{{\tt
  arXiv:1910.02080}}].

\bibitem{Ballesteros:2016xej}
G.~Ballesteros, J.~Redondo, A.~Ringwald, and C.~Tamarit, {\it {Standard
  Model\textemdash{}axion\textemdash{}seesaw\textemdash{}Higgs portal
  inflation. Five problems of particle physics and cosmology solved in one
  stroke}},  {\em JCAP} {\bf 08} (2017) 001,
  [\href{http://arxiv.org/abs/1610.01639}{{\tt arXiv:1610.01639}}].

\bibitem{Clarke:2015bea}
J.~D. Clarke and R.~R. Volkas, {\it {Technically natural nonsupersymmetric
  model of neutrino masses, baryogenesis, the strong CP problem, and dark
  matter}},  {\em Phys. Rev. D} {\bf 93} (2016), no.~3 035001,
  [\href{http://arxiv.org/abs/1509.07243}{{\tt arXiv:1509.07243}}].

\bibitem{Sopov:2022bog}
A.~H. Sopov and R.~R. Volkas, {\it {VISH\ensuremath{\nu}: solving five Standard
  Model shortcomings with a Poincar\'e-protected electroweak scale}},  {\em
  Phys. Dark Univ.} {\bf 42} (2023) 101381,
  [\href{http://arxiv.org/abs/2206.11598}{{\tt arXiv:2206.11598}}].

\bibitem{Dror:2019syi}
J.~A. Dror, T.~Hiramatsu, K.~Kohri, H.~Murayama, and G.~White, {\it {Testing
  the Seesaw Mechanism and Leptogenesis with Gravitational Waves}},  {\em Phys.
  Rev. Lett.} {\bf 124} (2020), no.~4 041804,
  [\href{http://arxiv.org/abs/1908.03227}{{\tt arXiv:1908.03227}}].

\bibitem{Blasi:2020wpy}
S.~Blasi, V.~Brdar, and K.~Schmitz, {\it {Fingerprint of low-scale leptogenesis
  in the primordial gravitational-wave spectrum}},  {\em Phys. Rev. Res.} {\bf
  2} (2020), no.~4 043321, [\href{http://arxiv.org/abs/2004.02889}{{\tt
  arXiv:2004.02889}}].

\bibitem{Fornal:2020esl}
B.~Fornal and B.~Shams Es~Haghi, {\it {Baryon and Lepton Number Violation from
  Gravitational Waves}},  {\em Phys. Rev. D} {\bf 102} (2020), no.~11 115037,
  [\href{http://arxiv.org/abs/2008.05111}{{\tt arXiv:2008.05111}}].

\bibitem{Samanta:2020cdk}
R.~Samanta and S.~Datta, {\it {Gravitational wave complementarity and impact of
  NANOGrav data on gravitational leptogenesis}},  {\em JHEP} {\bf 05} (2021)
  211, [\href{http://arxiv.org/abs/2009.13452}{{\tt arXiv:2009.13452}}].

\bibitem{Barman:2022yos}
B.~Barman, D.~Borah, A.~Dasgupta, and A.~Ghoshal, {\it {Probing High Scale
  Dirac Leptogenesis via Gravitational Waves from Domain Walls}},
  \href{http://arxiv.org/abs/2205.03422}{{\tt arXiv:2205.03422}}.

\bibitem{Visinelli:2009kt}
L.~Visinelli and P.~Gondolo, {\it {Axion cold dark matter in non-standard
  cosmologies}},  {\em Phys. Rev. D} {\bf 81} (2010) 063508,
  [\href{http://arxiv.org/abs/0912.0015}{{\tt arXiv:0912.0015}}].

\bibitem{Nelson:2018via}
A.~E. Nelson and H.~Xiao, {\it {Axion Cosmology with Early Matter Domination}},
   {\em Phys. Rev. D} {\bf 98} (2018), no.~6 063516,
  [\href{http://arxiv.org/abs/1807.07176}{{\tt arXiv:1807.07176}}].

\bibitem{Hawking:1974rv}
S.~W. Hawking, {\it {Black hole explosions}},  {\em Nature} {\bf 248} (1974)
  30--31.

\bibitem{Hawking:1975vcx}
S.~W. Hawking, {\it {Particle Creation by Black Holes}},  {\em Commun. Math.
  Phys.} {\bf 43} (1975) 199--220. [Erratum: Commun.Math.Phys. 46, 206 (1976)].

\bibitem{Carr:2020gox}
B.~Carr, K.~Kohri, Y.~Sendouda, and J.~Yokoyama, {\it {Constraints on
  primordial black holes}},  {\em Rept. Prog. Phys.} {\bf 84} (2021), no.~11
  116902, [\href{http://arxiv.org/abs/2002.12778}{{\tt arXiv:2002.12778}}].

\bibitem{Papanikolaou:2020qtd}
T.~Papanikolaou, V.~Vennin, and D.~Langlois, {\it {Gravitational waves from a
  universe filled with primordial black holes}},  {\em JCAP} {\bf 03} (2021)
  053, [\href{http://arxiv.org/abs/2010.11573}{{\tt arXiv:2010.11573}}].

\bibitem{Domenech:2020ssp}
G.~Dom\`enech, C.~Lin, and M.~Sasaki, {\it {Gravitational wave constraints on
  the primordial black hole dominated early universe}},  {\em JCAP} {\bf 04}
  (2021) 062, [\href{http://arxiv.org/abs/2012.08151}{{\tt arXiv:2012.08151}}].
  [Erratum: JCAP 11, E01 (2021)].

\bibitem{Domenech:2021wkk}
G.~Dom\`enech, V.~Takhistov, and M.~Sasaki, {\it {Exploring evaporating
  primordial black holes with gravitational waves}},  {\em Phys. Lett. B} {\bf
  823} (2021) 136722, [\href{http://arxiv.org/abs/2105.06816}{{\tt
  arXiv:2105.06816}}].

\bibitem{Domenech:2021ztg}
G.~Dom\`enech, {\it {Scalar Induced Gravitational Waves Review}},  {\em
  Universe} {\bf 7} (2021), no.~11 398,
  [\href{http://arxiv.org/abs/2109.01398}{{\tt arXiv:2109.01398}}].

\bibitem{Papanikolaou:2022chm}
T.~Papanikolaou, {\it {Gravitational waves induced from primordial black hole
  fluctuations: the~effect of an extended mass function}},  {\em JCAP} {\bf 10}
  (2022) 089, [\href{http://arxiv.org/abs/2207.11041}{{\tt arXiv:2207.11041}}].

\bibitem{Bernal:2021yyb}
N.~Bernal, F.~Hajkarim, and Y.~Xu, {\it {Axion Dark Matter in the Time of
  Primordial Black Holes}},  {\em Phys. Rev. D} {\bf 104} (2021) 075007,
  [\href{http://arxiv.org/abs/2107.13575}{{\tt arXiv:2107.13575}}].

\bibitem{Mazde:2022sdx}
K.~Mazde and L.~Visinelli, {\it {The interplay between the dark matter axion
  and primordial black holes}},  {\em JCAP} {\bf 01} (2023) 021,
  [\href{http://arxiv.org/abs/2209.14307}{{\tt arXiv:2209.14307}}].

\bibitem{Gondolo:2020uqv}
P.~Gondolo, P.~Sandick, and B.~Shams Es~Haghi, {\it {Effects of primordial
  black holes on dark matter models}},  {\em Phys. Rev. D} {\bf 102} (2020),
  no.~9 095018, [\href{http://arxiv.org/abs/2009.02424}{{\tt
  arXiv:2009.02424}}].

\bibitem{Bernal:2020bjf}
N.~Bernal and O.~Zapata, {\it {Dark Matter in the Time of Primordial Black
  Holes}},  {\em JCAP} {\bf 03} (2021) 015,
  [\href{http://arxiv.org/abs/2011.12306}{{\tt arXiv:2011.12306}}].

\bibitem{Green:1999yh}
A.~M. Green, {\it {Supersymmetry and primordial black hole abundance
  constraints}},  {\em Phys. Rev. D} {\bf 60} (1999) 063516,
  [\href{http://arxiv.org/abs/astro-ph/9903484}{{\tt astro-ph/9903484}}].

\bibitem{Khlopov:2004tn}
M.~Y. Khlopov, A.~Barrau, and J.~Grain, {\it {Gravitino production by
  primordial black hole evaporation and constraints on the inhomogeneity of the
  early universe}},  {\em Class. Quant. Grav.} {\bf 23} (2006) 1875--1882,
  [\href{http://arxiv.org/abs/astro-ph/0406621}{{\tt astro-ph/0406621}}].

\bibitem{Dai:2009hx}
D.-C. Dai, K.~Freese, and D.~Stojkovic, {\it {Constraints on dark matter
  particles charged under a hidden gauge group from primordial black holes}},
  {\em JCAP} {\bf 06} (2009) 023, [\href{http://arxiv.org/abs/0904.3331}{{\tt
  arXiv:0904.3331}}].

\bibitem{Allahverdi:2017sks}
R.~Allahverdi, J.~Dent, and J.~Osinski, {\it {Nonthermal production of dark
  matter from primordial black holes}},  {\em Phys. Rev. D} {\bf 97} (2018),
  no.~5 055013, [\href{http://arxiv.org/abs/1711.10511}{{\tt
  arXiv:1711.10511}}].

\bibitem{Lennon:2017tqq}
O.~Lennon, J.~March-Russell, R.~Petrossian-Byrne, and H.~Tillim, {\it {Black
  Hole Genesis of Dark Matter}},  {\em JCAP} {\bf 04} (2018) 009,
  [\href{http://arxiv.org/abs/1712.07664}{{\tt arXiv:1712.07664}}].

\bibitem{Hooper:2019gtx}
D.~Hooper, G.~Krnjaic, and S.~D. McDermott, {\it {Dark Radiation and Superheavy
  Dark Matter from Black Hole Domination}},  {\em JHEP} {\bf 08} (2019) 001,
  [\href{http://arxiv.org/abs/1905.01301}{{\tt arXiv:1905.01301}}].

\bibitem{Sandick:2021gew}
P.~Sandick, B.~S. Es~Haghi, and K.~Sinha, {\it {Asymmetric reheating by
  primordial black holes}},  {\em Phys. Rev. D} {\bf 104} (2021), no.~8 083523,
  [\href{http://arxiv.org/abs/2108.08329}{{\tt arXiv:2108.08329}}].

\bibitem{Fujita:2014hha}
T.~Fujita, M.~Kawasaki, K.~Harigaya, and R.~Matsuda, {\it {Baryon asymmetry,
  dark matter, and density perturbation from primordial black holes}},  {\em
  Phys. Rev. D} {\bf 89} (2014), no.~10 103501,
  [\href{http://arxiv.org/abs/1401.1909}{{\tt arXiv:1401.1909}}].

\bibitem{Datta:2020bht}
S.~Datta, A.~Ghosal, and R.~Samanta, {\it {Baryogenesis from ultralight
  primordial black holes and strong gravitational waves from cosmic strings}},
  {\em JCAP} {\bf 08} (2021) 021, [\href{http://arxiv.org/abs/2012.14981}{{\tt
  arXiv:2012.14981}}].

\bibitem{JyotiDas:2021shi}
S.~Jyoti~Das, D.~Mahanta, and D.~Borah, {\it {Low scale leptogenesis and dark
  matter in the presence of primordial black holes}},
  \href{http://arxiv.org/abs/2104.14496}{{\tt arXiv:2104.14496}}.

\bibitem{Barman:2021ost}
B.~Barman, D.~Borah, S.~J. Das, and R.~Roshan, {\it {Non-thermal origin of
  asymmetric dark matter from inflaton and primordial black holes}},  {\em
  JCAP} {\bf 03} (2022), no.~03 031,
  [\href{http://arxiv.org/abs/2111.08034}{{\tt arXiv:2111.08034}}].

\bibitem{Barman:2022gjo}
B.~Barman, D.~Borah, S.~Das~Jyoti, and R.~Roshan, {\it {Cogenesis of Baryon
  Asymmetry and Gravitational Dark Matter from PBH}},
  \href{http://arxiv.org/abs/2204.10339}{{\tt arXiv:2204.10339}}.

\bibitem{Barman:2022pdo}
B.~Barman, D.~Borah, S.~Jyoti~Das, and R.~Roshan, {\it {Gravitational wave
  signatures of a PBH-generated baryon-dark matter coincidence}},  {\em Phys.
  Rev. D} {\bf 107} (2023), no.~9 095002,
  [\href{http://arxiv.org/abs/2212.00052}{{\tt arXiv:2212.00052}}].

\bibitem{Cheek:2021odj}
A.~Cheek, L.~Heurtier, Y.~F. Perez-Gonzalez, and J.~Turner, {\it {Primordial
  Black Hole Evaporation and Dark Matter Production: I. Solely Hawking
  radiation}},  \href{http://arxiv.org/abs/2107.00013}{{\tt arXiv:2107.00013}}.

\bibitem{Chaudhuri:2023aiv}
A.~Chaudhuri, B.~Coleppa, and K.~Loho, {\it {Dark matter production from two
  evaporating PBH distributions}},  {\em Phys. Rev. D} {\bf 108} (2023), no.~3
  035040, [\href{http://arxiv.org/abs/2301.08588}{{\tt arXiv:2301.08588}}].

\bibitem{Borah:2024lml}
D.~Borah, S.~Jyoti~Das, and I.~Saha, {\it {Dark matter from phase transition
  generated PBH evaporation with gravitational waves signatures}},
  \href{http://arxiv.org/abs/2401.12282}{{\tt arXiv:2401.12282}}.

\bibitem{Carr:1976zz}
B.~J. Carr, {\it {Some cosmological consequences of primordial black-hole
  evaporations}},  {\em Astrophys. J.} {\bf 206} (1976) 8--25.

\bibitem{Baumann:2007yr}
D.~Baumann, P.~J. Steinhardt, and N.~Turok, {\it {Primordial Black Hole
  Baryogenesis}},  \href{http://arxiv.org/abs/hep-th/0703250}{{\tt
  hep-th/0703250}}.

\bibitem{Hook:2014mla}
A.~Hook, {\it {Baryogenesis from Hawking Radiation}},  {\em Phys. Rev. D} {\bf
  90} (2014), no.~8 083535, [\href{http://arxiv.org/abs/1404.0113}{{\tt
  arXiv:1404.0113}}].

\bibitem{Hamada:2016jnq}
Y.~Hamada and S.~Iso, {\it {Baryon asymmetry from primordial black holes}},
  {\em PTEP} {\bf 2017} (2017), no.~3 033B02,
  [\href{http://arxiv.org/abs/1610.02586}{{\tt arXiv:1610.02586}}].

\bibitem{Morrison:2018xla}
L.~Morrison, S.~Profumo, and Y.~Yu, {\it {Melanopogenesis: Dark Matter of
  (almost) any Mass and Baryonic Matter from the Evaporation of Primordial
  Black Holes weighing a Ton (or less)}},  {\em JCAP} {\bf 05} (2019) 005,
  [\href{http://arxiv.org/abs/1812.10606}{{\tt arXiv:1812.10606}}].

\bibitem{Hooper:2020otu}
D.~Hooper and G.~Krnjaic, {\it {GUT Baryogenesis With Primordial Black Holes}},
   {\em Phys. Rev. D} {\bf 103} (2021), no.~4 043504,
  [\href{http://arxiv.org/abs/2010.01134}{{\tt arXiv:2010.01134}}].

\bibitem{Perez-Gonzalez:2020vnz}
Y.~F. Perez-Gonzalez and J.~Turner, {\it {Assessing the tension between a black
  hole dominated early universe and leptogenesis}},
  \href{http://arxiv.org/abs/2010.03565}{{\tt arXiv:2010.03565}}.

\bibitem{Smyth:2021lkn}
N.~Smyth, L.~Santos-Olmsted, and S.~Profumo, {\it {Gravitational Baryogenesis
  and Dark Matter from Light Black Holes}},
  \href{http://arxiv.org/abs/2110.14660}{{\tt arXiv:2110.14660}}.

\bibitem{Bernal:2022pue}
N.~Bernal, C.~S. Fong, Y.~F. Perez-Gonzalez, and J.~Turner, {\it {Rescuing
  High-Scale Leptogenesis using Primordial Black Holes}},
  \href{http://arxiv.org/abs/2203.08823}{{\tt arXiv:2203.08823}}.

\bibitem{Ambrosone:2021lsx}
A.~Ambrosone, R.~Calabrese, D.~F.~G. Fiorillo, G.~Miele, and S.~Morisi, {\it
  {Towards baryogenesis via absorption from primordial black holes}},  {\em
  Phys. Rev. D} {\bf 105} (2022), no.~4 045001,
  [\href{http://arxiv.org/abs/2106.11980}{{\tt arXiv:2106.11980}}].

\bibitem{Franciolini:2021nvv}
G.~Franciolini, {\em {Primordial Black Holes: from Theory to Gravitational Wave
  Observations}}.
\newblock PhD thesis, Geneva U., Dept. Theor. Phys., 2021.
\newblock \href{http://arxiv.org/abs/2110.06815}{{\tt arXiv:2110.06815}}.

\bibitem{Yoo:2022mzl}
C.-M. Yoo, {\it {The Basics of Primordial Black Hole Formation and Abundance
  Estimation}},  {\em Galaxies} {\bf 10} (2022), no.~6 112,
  [\href{http://arxiv.org/abs/2211.13512}{{\tt arXiv:2211.13512}}].

\bibitem{Bhattacharya:2023ztw}
S.~Bhattacharya, {\it {Primordial Black Hole Formation in Non-Standard
  Post-Inflationary Epochs}},  {\em Galaxies} {\bf 11} (2023), no.~1 35,
  [\href{http://arxiv.org/abs/2302.12690}{{\tt arXiv:2302.12690}}].

\bibitem{Ferrer:2018uiu}
F.~Ferrer, E.~Masso, G.~Panico, O.~Pujolas, and F.~Rompineve, {\it {Primordial
  Black Holes from the QCD axion}},  {\em Phys. Rev. Lett.} {\bf 122} (2019),
  no.~10 101301, [\href{http://arxiv.org/abs/1807.01707}{{\tt
  arXiv:1807.01707}}].

\bibitem{Minkowski:1977sc}
P.~Minkowski, {\it {$\mu \to e\gamma$ at a Rate of One Out of $10^{9}$ Muon
  Decays?}},  {\em Phys. Lett. B} {\bf 67} (1977) 421--428.

\bibitem{GellMann:1980vs}
M.~Gell-Mann, P.~Ramond, and R.~Slansky, {\it {Complex Spinors and Unified
  Theories}},  {\em Conf. Proc. C} {\bf 790927} (1979) 315--321,
  [\href{http://arxiv.org/abs/1306.4669}{{\tt arXiv:1306.4669}}].

\bibitem{Mohapatra:1979ia}
R.~N. Mohapatra and G.~Senjanovic, {\it {Neutrino Mass and Spontaneous Parity
  Nonconservation}},  {\em Phys. Rev. Lett.} {\bf 44} (1980) 912.

\bibitem{Yanagida:1980xy}
T.~Yanagida, {\it {Horizontal Symmetry and Masses of Neutrinos}},  {\em Prog.
  Theor. Phys.} {\bf 64} (1980) 1103.

\bibitem{Schechter:1980gr}
J.~Schechter and J.~Valle, {\it {Neutrino Masses in SU(2) x U(1) Theories}},
  {\em Phys. Rev. D} {\bf 22} (1980) 2227.

\bibitem{DiLuzio:2020wdo}
L.~Di~Luzio, M.~Giannotti, E.~Nardi, and L.~Visinelli, {\it {The landscape of
  QCD axion models}},  {\em Phys. Rept.} {\bf 870} (2020) 1--117,
  [\href{http://arxiv.org/abs/2003.01100}{{\tt arXiv:2003.01100}}].

\bibitem{Masina:2020xhk}
I.~Masina, {\it {Dark matter and dark radiation from evaporating primordial
  black holes}},  {\em Eur. Phys. J. Plus} {\bf 135} (2020), no.~7 552,
  [\href{http://arxiv.org/abs/2004.04740}{{\tt arXiv:2004.04740}}].

\bibitem{Pascoli:2006ci}
S.~Pascoli, S.~T. Petcov, and A.~Riotto, {\it {Leptogenesis and Low Energy CP
  Violation in Neutrino Physics}},  {\em Nucl. Phys. B} {\bf 774} (2007) 1--52,
  [\href{http://arxiv.org/abs/hep-ph/0611338}{{\tt hep-ph/0611338}}].

\bibitem{Buchmuller:2004nz}
W.~Buchmuller, P.~Di~Bari, and M.~Plumacher, {\it {Leptogenesis for
  pedestrians}},  {\em Annals Phys.} {\bf 315} (2005) 305--351,
  [\href{http://arxiv.org/abs/hep-ph/0401240}{{\tt hep-ph/0401240}}].

\bibitem{Davidson:2008bu}
S.~Davidson, E.~Nardi, and Y.~Nir, {\it {Leptogenesis}},  {\em Phys. Rept.}
  {\bf 466} (2008) 105--177, [\href{http://arxiv.org/abs/0802.2962}{{\tt
  arXiv:0802.2962}}].

\bibitem{Borah:2022iym}
D.~Borah, S.~Jyoti~Das, R.~Samanta, and F.~R. Urban, {\it {PBH-infused seesaw
  origin of matter and unique gravitational waves}},  {\em JHEP} {\bf 03}
  (2023) 127, [\href{http://arxiv.org/abs/2211.15726}{{\tt arXiv:2211.15726}}].

\bibitem{Pilaftsis:2003gt}
A.~Pilaftsis and T.~E.~J. Underwood, {\it {Resonant leptogenesis}},  {\em Nucl.
  Phys. B} {\bf 692} (2004) 303--345,
  [\href{http://arxiv.org/abs/hep-ph/0309342}{{\tt hep-ph/0309342}}].

\bibitem{Dev:2017wwc}
P.~S.~B. Dev, M.~Garny, J.~Klaric, P.~Millington, and D.~Teresi, {\it {Resonant
  enhancement in leptogenesis}},  {\em Int. J. Mod. Phys.} {\bf A33} (2018)
  1842003, [\href{http://arxiv.org/abs/1711.02863}{{\tt arXiv:1711.02863}}].

\bibitem{Abazajian:2019eic}
K.~Abazajian et~al., {\it {CMB-S4 Science Case, Reference Design, and Project
  Plan}},  \href{http://arxiv.org/abs/1907.04473}{{\tt arXiv:1907.04473}}.

\bibitem{Inomata:2019ivs}
K.~Inomata, K.~Kohri, T.~Nakama, and T.~Terada, {\it {Enhancement of
  Gravitational Waves Induced by Scalar Perturbations due to a Sudden
  Transition from an Early Matter Era to the Radiation Era}},  {\em Phys. Rev.
  D} {\bf 100} (2019), no.~4 043532,
  [\href{http://arxiv.org/abs/1904.12879}{{\tt arXiv:1904.12879}}].

\bibitem{LIGOScientific:2014pky}
{\bf LIGO Scientific} Collaboration, J.~Aasi et~al., {\it {Advanced LIGO}},
  {\em Class. Quant. Grav.} {\bf 32} (2015) 074001,
  [\href{http://arxiv.org/abs/1411.4547}{{\tt arXiv:1411.4547}}].

\bibitem{2017arXiv170200786A}
{\bf LISA} Collaboration, P.~Amaro-Seoane~et al, {\it {Laser Interferometer
  Space Antenna}},  {\em arXiv e-prints} (Feb., 2017) arXiv:1702.00786,
  [\href{http://arxiv.org/abs/1702.00786}{{\tt arXiv:1702.00786}}].

\bibitem{Kawamura:2006up}
S.~Kawamura et~al., {\it {The Japanese space gravitational wave antenna
  DECIGO}},  {\em Class. Quant. Grav.} {\bf 23} (2006) S125--S132.

\bibitem{LIGOScientific:2016wof}
{\bf LIGO Scientific} Collaboration, B.~P. Abbott et~al., {\it {Exploring the
  Sensitivity of Next Generation Gravitational Wave Detectors}},  {\em Class.
  Quant. Grav.} {\bf 34} (2017), no.~4 044001,
  [\href{http://arxiv.org/abs/1607.08697}{{\tt arXiv:1607.08697}}].

\bibitem{Sato:2017dkf}
S.~Sato et~al., {\it {The status of DECIGO}},  {\em J. Phys. Conf. Ser.} {\bf
  840} (2017), no.~1 012010.

\bibitem{Ishikawa:2020hlo}
T.~Ishikawa et~al., {\it {Improvement of the target sensitivity in DECIGO by
  optimizing its parameters for quantum noise including the effect of
  diffraction loss}},  {\em Galaxies} {\bf 9} (2021), no.~1 14,
  [\href{http://arxiv.org/abs/2012.11859}{{\tt arXiv:2012.11859}}].

\bibitem{Baeza-Ballesteros:2023say}
J.~Baeza-Ballesteros, E.~J. Copeland, D.~G. Figueroa, and J.~Lizarraga, {\it
  {Gravitational Wave Emission from a Cosmic String Loop, I: Global Case}},
  \href{http://arxiv.org/abs/2308.08456}{{\tt arXiv:2308.08456}}.

\bibitem{Gorghetto:2021fsn}
M.~Gorghetto, E.~Hardy, and H.~Nicolaescu, {\it {Observing invisible axions
  with gravitational waves}},  {\em JCAP} {\bf 06} (2021) 034,
  [\href{http://arxiv.org/abs/2101.11007}{{\tt arXiv:2101.11007}}].

\bibitem{Vilenkin:2018zol}
A.~Vilenkin, Y.~Levin, and A.~Gruzinov, {\it {Cosmic strings and primordial
  black holes}},  {\em JCAP} {\bf 11} (2018) 008,
  [\href{http://arxiv.org/abs/1808.00670}{{\tt arXiv:1808.00670}}].

\bibitem{Srednicki:1985xd}
M.~Srednicki, {\it {Axion Couplings to Matter. 1. CP Conserving Parts}},  {\em
  Nucl. Phys. B} {\bf 260} (1985) 689--700.

\bibitem{CAST:2007jps}
{\bf CAST} Collaboration, S.~Andriamonje et~al., {\it {An Improved limit on the
  axion-photon coupling from the CAST experiment}},  {\em JCAP} {\bf 04} (2007)
  010, [\href{http://arxiv.org/abs/hep-ex/0702006}{{\tt hep-ex/0702006}}].

\bibitem{CAST:2017uph}
{\bf CAST} Collaboration, V.~Anastassopoulos et~al., {\it {New CAST Limit on
  the Axion-Photon Interaction}},  {\em Nature Phys.} {\bf 13} (2017) 584--590,
  [\href{http://arxiv.org/abs/1705.02290}{{\tt arXiv:1705.02290}}].

\bibitem{PhysRevLett.60.1797}
M.~S. Turner, {\it Axions from sn1987a},  {\em Phys. Rev. Lett.} {\bf 60} (May,
  1988) 1797--1800.

\bibitem{PhysRevD.39.1020}
A.~Burrows, M.~S. Turner, and R.~P. Brinkmann, {\it Axions and sn 1987a},  {\em
  Phys. Rev. D} {\bf 39} (Feb, 1989) 1020--1028.

\bibitem{PhysRevLett.60.1793}
G.~Raffelt and D.~Seckel, {\it Bounds on exotic-particle interactions from
  sn1987a},  {\em Phys. Rev. Lett.} {\bf 60} (May, 1988) 1793--1796.

\bibitem{Fermi-LAT:2016nkz}
{\bf Fermi-LAT} Collaboration, M.~Ajello et~al., {\it {Search for Spectral
  Irregularities due to Photon\textendash{}Axionlike-Particle Oscillations with
  the Fermi Large Area Telescope}},  {\em Phys. Rev. Lett.} {\bf 116} (2016),
  no.~16 161101, [\href{http://arxiv.org/abs/1603.06978}{{\tt
  arXiv:1603.06978}}].

\bibitem{ADMX:2006kgb}
{\bf ADMX} Collaboration, L.~D. Duffy, P.~Sikivie, D.~B. Tanner, S.~J.
  Asztalos, C.~Hagmann, D.~Kinion, L.~J. Rosenberg, K.~van Bibber, D.~B. Yu,
  and R.~F. Bradley, {\it {A high resolution search for dark-matter axions}},
  {\em Phys. Rev. D} {\bf 74} (2006) 012006,
  [\href{http://arxiv.org/abs/astro-ph/0603108}{{\tt astro-ph/0603108}}].

\bibitem{Stern:2016bbw}
I.~Stern, {\it {ADMX Status}},  {\em PoS} {\bf ICHEP2016} (2016) 198,
  [\href{http://arxiv.org/abs/1612.08296}{{\tt arXiv:1612.08296}}].

\bibitem{ADMX:2019uok}
{\bf ADMX} Collaboration, T.~Braine et~al., {\it {Extended Search for the
  Invisible Axion with the Axion Dark Matter Experiment}},  {\em Phys. Rev.
  Lett.} {\bf 124} (2020), no.~10 101303,
  [\href{http://arxiv.org/abs/1910.08638}{{\tt arXiv:1910.08638}}].

\bibitem{Budker:2013hfa}
D.~Budker, P.~W. Graham, M.~Ledbetter, S.~Rajendran, and A.~Sushkov, {\it
  {Proposal for a Cosmic Axion Spin Precession Experiment (CASPEr)}},  {\em
  Phys. Rev. X} {\bf 4} (2014), no.~2 021030,
  [\href{http://arxiv.org/abs/1306.6089}{{\tt arXiv:1306.6089}}].

\bibitem{Alesini:2017ifp}
D.~Alesini, D.~Babusci, D.~Di~Gioacchino, C.~Gatti, G.~Lamanna, and C.~Ligi,
  {\it {The KLASH Proposal}},  \href{http://arxiv.org/abs/1707.06010}{{\tt
  arXiv:1707.06010}}.

\bibitem{Alesini:2019nzq}
D.~Alesini et~al., {\it {KLASH Conceptual Design Report}},
  \href{http://arxiv.org/abs/1911.02427}{{\tt arXiv:1911.02427}}.

\bibitem{Alesini:2023qed}
D.~Alesini et~al., {\it {The future search for low-frequency axions and new
  physics with the FLASH resonant cavity experiment at Frascati National
  Laboratories}},  {\em Phys. Dark Univ.} {\bf 42} (2023) 101370,
  [\href{http://arxiv.org/abs/2309.00351}{{\tt arXiv:2309.00351}}].

\bibitem{Kahn:2016aff}
Y.~Kahn, B.~R. Safdi, and J.~Thaler, {\it {Broadband and Resonant Approaches to
  Axion Dark Matter Detection}},  {\em Phys. Rev. Lett.} {\bf 117} (2016),
  no.~14 141801, [\href{http://arxiv.org/abs/1602.01086}{{\tt
  arXiv:1602.01086}}].

\bibitem{Ouellet:2018beu}
J.~L. Ouellet et~al., {\it {First Results from ABRACADABRA-10 cm: A Search for
  Sub-$\mu$eV Axion Dark Matter}},  {\em Phys. Rev. Lett.} {\bf 122} (2019),
  no.~12 121802, [\href{http://arxiv.org/abs/1810.12257}{{\tt
  arXiv:1810.12257}}].

\bibitem{Lee:2020cfj}
S.~Lee, S.~Ahn, J.~Choi, B.~R. Ko, and Y.~K. Semertzidis, {\it {Axion Dark
  Matter Search around 6.7 $\mu$eV}},  {\em Phys. Rev. Lett.} {\bf 124} (2020),
  no.~10 101802, [\href{http://arxiv.org/abs/2001.05102}{{\tt
  arXiv:2001.05102}}].

\bibitem{Semertzidis:2019gkj}
Y.~K. Semertzidis et~al., {\it {Axion Dark Matter Research with IBS/CAPP}},
  \href{http://arxiv.org/abs/1910.11591}{{\tt arXiv:1910.11591}}.

\bibitem{Caldwell:2016dcw}
{\bf MADMAX Working Group} Collaboration, A.~Caldwell, G.~Dvali, B.~Majorovits,
  A.~Millar, G.~Raffelt, J.~Redondo, O.~Reimann, F.~Simon, and F.~Steffen, {\it
  {Dielectric Haloscopes: A New Way to Detect Axion Dark Matter}},  {\em Phys.
  Rev. Lett.} {\bf 118} (2017), no.~9 091801,
  [\href{http://arxiv.org/abs/1611.05865}{{\tt arXiv:1611.05865}}].

\bibitem{Vogel:2013bta}
J.~K. Vogel et~al., {\it {IAXO - The International Axion Observatory}},  in
  {\em {8th Patras Workshop on Axions, WIMPs and WISPs}}, 2, 2013.
\newblock \href{http://arxiv.org/abs/1302.3273}{{\tt arXiv:1302.3273}}.

\bibitem{IAXO:2019mpb}
{\bf IAXO} Collaboration, E.~Armengaud et~al., {\it {Physics potential of the
  International Axion Observatory (IAXO)}},  {\em JCAP} {\bf 06} (2019) 047,
  [\href{http://arxiv.org/abs/1904.09155}{{\tt arXiv:1904.09155}}].

\bibitem{Meyer:2016wrm}
M.~Meyer, M.~Giannotti, A.~Mirizzi, J.~Conrad, and M.~A. S\'anchez-Conde, {\it
  {Fermi Large Area Telescope as a Galactic Supernovae Axionscope}},  {\em
  Phys. Rev. Lett.} {\bf 118} (2017), no.~1 011103,
  [\href{http://arxiv.org/abs/1609.02350}{{\tt arXiv:1609.02350}}].

\bibitem{Cardoso:2018tly}
V.~Cardoso, O.~J.~C. Dias, G.~S. Hartnett, M.~Middleton, P.~Pani, and J.~E.
  Santos, {\it {Constraining the mass of dark photons and axion-like particles
  through black-hole superradiance}},  {\em JCAP} {\bf 03} (2018) 043,
  [\href{http://arxiv.org/abs/1801.01420}{{\tt arXiv:1801.01420}}].

\bibitem{TASTE:2017pdv}
{\bf TASTE} Collaboration, V.~Anastassopoulos et~al., {\it {Towards a
  medium-scale axion helioscope and haloscope}},  {\em JINST} {\bf 12} (2017),
  no.~11 P11019, [\href{http://arxiv.org/abs/1706.09378}{{\tt
  arXiv:1706.09378}}].

\bibitem{Co:2017mop}
R.~T. Co, L.~J. Hall, and K.~Harigaya, {\it {QCD Axion Dark Matter with a Small
  Decay Constant}},  {\em Phys. Rev. Lett.} {\bf 120} (2018), no.~21 211602,
  [\href{http://arxiv.org/abs/1711.10486}{{\tt arXiv:1711.10486}}].

\bibitem{Co:2020dya}
R.~T. Co, L.~J. Hall, K.~Harigaya, K.~A. Olive, and S.~Verner, {\it {Axion
  Kinetic Misalignment and Parametric Resonance from Inflation}},  {\em JCAP}
  {\bf 08} (2020) 036, [\href{http://arxiv.org/abs/2004.00629}{{\tt
  arXiv:2004.00629}}].

\bibitem{Ramazanov:2022kbd}
S.~Ramazanov and R.~Samanta, {\it {Heating up Peccei-Quinn scale}},  {\em JCAP}
  {\bf 05} (2023) 048, [\href{http://arxiv.org/abs/2210.08407}{{\tt
  arXiv:2210.08407}}].

\bibitem{Dodelson:1993je}
S.~Dodelson and L.~M. Widrow, {\it {Sterile-neutrinos as dark matter}},  {\em
  Phys. Rev. Lett.} {\bf 72} (1994) 17--20,
  [\href{http://arxiv.org/abs/hep-ph/9303287}{{\tt hep-ph/9303287}}].

\bibitem{Drewes:2016upu}
M.~Drewes et~al., {\it {A White Paper on keV Sterile Neutrino Dark Matter}},
  {\em JCAP} {\bf 01} (2017) 025, [\href{http://arxiv.org/abs/1602.04816}{{\tt
  arXiv:1602.04816}}].

\bibitem{Ghosh:2023tyz}
D.~K. Ghosh, A.~Ghoshal, and S.~Jeesun, {\it {Axion-like particle (ALP) portal
  freeze-in dark matter confronting ALP search experiments}},  {\em JHEP} {\bf
  01} (2024) 026, [\href{http://arxiv.org/abs/2305.09188}{{\tt
  arXiv:2305.09188}}].

\bibitem{Bharucha:2022lty}
A.~Bharucha, F.~Br\"ummer, N.~Desai, and S.~Mutzel, {\it {Axion-like particles
  as mediators for dark matter: beyond freeze-out}},  {\em JHEP} {\bf 02}
  (2023) 141, [\href{http://arxiv.org/abs/2209.03932}{{\tt arXiv:2209.03932}}].

\bibitem{Fitzpatrick:2023xks}
P.~J. Fitzpatrick, Y.~Hochberg, E.~Kuflik, R.~Ovadia, and Y.~Soreq, {\it {Dark
  matter through the axion-gluon portal}},  {\em Phys. Rev. D} {\bf 108}
  (2023), no.~7 075003, [\href{http://arxiv.org/abs/2306.03128}{{\tt
  arXiv:2306.03128}}].

\bibitem{Borah:2020wyc}
D.~Borah, S.~Jyoti~Das, and A.~K. Saha, {\it {Cosmic inflation in minimal
  $U(1)_{B-L}$ model: implications for (non) thermal dark matter and
  leptogenesis}},  {\em Eur. Phys. J. C} {\bf 81} (2021), no.~2 169,
  [\href{http://arxiv.org/abs/2005.11328}{{\tt arXiv:2005.11328}}].

\bibitem{Datta:2021elq}
A.~Datta, R.~Roshan, and A.~Sil, {\it {Imprint of the Seesaw Mechanism on
  Feebly Interacting Dark Matter and the Baryon Asymmetry}},  {\em Phys. Rev.
  Lett.} {\bf 127} (2021), no.~23 231801,
  [\href{http://arxiv.org/abs/2104.02030}{{\tt arXiv:2104.02030}}].

\bibitem{Griest:1989wd}
K.~Griest and M.~Kamionkowski, {\it {Unitarity Limits on the Mass and Radius of
  Dark Matter Particles}},  {\em Phys. Rev. Lett.} {\bf 64} (1990) 615.

\bibitem{Giudice:2000ex}
G.~F. Giudice, E.~W. Kolb, and A.~Riotto, {\it {Largest temperature of the
  radiation era and its cosmological implications}},  {\em Phys. Rev. D} {\bf
  64} (2001) 023508, [\href{http://arxiv.org/abs/hep-ph/0005123}{{\tt
  hep-ph/0005123}}].

\end{thebibliography}\endgroup
\end{document}